# Phonon anharmonicity: a pertinent review of recent progress and perspective

Bin Wei[1,2], Qiyang Sun[3], Chen Li[3,4], and Jiawang Hong[2]*

[1]Henan Key Laboratory of Materials on Deep-Earth Engineering, School of Materials Science and Engineering, Henan Polytechnic University, Jiaozuo 454000, China

[2]School of Aerospace Engineering, Beijing Institute of Technology, Beijing 100081, China

[3]Department of Mechanical Engineering, University of California, Riverside, Riverside, CA 92521, USA

[4]Materials Science and Engineering, University of California, Riverside, Riverside, CA 92521, USA

* Corresponding author (e-mail: hongjw@bit.edu.cn)

Anharmonic lattice vibrations govern the thermal dynamics in materials and present how the atoms interact and how they conduct heat. An in-depth understanding of the microscopic mechanism of phonon anharmonicity in condensed systems is critical for developing better functional and energy materials. In recent years, a variety of novel behaviors in condense matters are driven by phonon anharmonic effects in some way or another, such as soft mode phase transition, negative thermal expansion (NTE), multiferroicity, ultralow thermal conductivity ($\kappa$) or high thermal resistance, and high-temperature superconductivity, etc. All these properties have endowed anharmonicity with many promising applications and provided remarkable opportunities for developing 'anharmonicity engineering' — regulating heat transport towards excellent performance in materials. In this work, we review the recent development of the study on phonon anharmonic effect and summarize its origination, influence and mechanism, research methods, and applications. Besides, the remaining challenges, future trends, and prospects of phonon anharmonicity are also put forward.

## Contents

# 1. Introduction

1.1 Background

In a solid system, almost all atoms (ion cores) vibrate at their equilibrium positions. In such context, many of the thermodynamics properties have been widely studied up to the present. However, a detailed description of these properties requires an understanding of the interactions between the normal phonon modes of these vibrations, especially for systems incorporating light atoms, with weak bonding, or at high temperatures [1]. Due to the limitations in the theory of harmonic or quasi-harmonic theory (i.e., the thermal effects depend only on the volume change) [2-4], the lattice anharmonicity has begun to emerge as a subject of considerable experimental and theoretical studies over the last century.

In classical mechanics, anharmonicity is defined as the nonlinear relationship between any oscillator with the generalized forces and the generalized displacement. However, such definition is not entirely accurate in vibrational thermodynamics due to the involvement of quasi-harmonic effects. Therefore, anharmonicity here is more restrictively defined as the effect that occurs beyond the harmonic and quasi-harmonic limit [5]. In this review, only the "pure" (phonon-phonon interactions) anharmonicity is concerned, and the harmonic and quasi-harmonic effects will not be discussed further. Interested readers can find elaborations extensively elsewhere [6,7].

In thermodynamics, anharmonicity was first introduced with the temperature-dependent lattice constant by Mie [8] and Grüneisen [9] in the early 1900s and was later studied by Born and co-workers [10-13]; however, the quasi-harmonic approximation was still involved when dealing with their formulations. The influence of anharmonicity on the caloric equations of state, carried out beyond the quasi-harmonic approximation, was first investigated by Born and Brody [14-16]. Later, Leibfried [17] modified the expressions for the free energy containing both thermal and caloric equations of state with anharmonic effects. In the 1930s, Blackman contributed the first papers on anharmonic effects in the absorption of infrared radiation by ionic crystals [18,19]. Henceforth, in some simple systems, many experimental and theoretical studies were performed in great detail to investigate the lattice anharmonicity [20-28], which remarkably developed and promoted the research on anharmonicity in the mid-1990s. Among these, two review papers help us to understand the bud of anharmonicity historically: one is the lattice dynamics of metals reviewed by Joshi and Rajagopal [20]; the other one is the anharmonicity of ionic crystals reviewed by Cochran and Cowley [28]. In the 1960s, the developments of anharmonic effects in lattice dynamics were systematically reviewed by Leibfried and Ludwig [15]. Since then, with the maturity of the anharmonic theories and the development of advanced measurement and computational technologies [5,29,30], the activity within the area of anharmonicity has become exciting. Numerous novel phenomena, such as negative thermal expansion (NTE) [31], ultralow thermal conductivity [32], and multiferroicity [33], have been deeply understood in the context of anharmonicity over the past decades.

In recent years the study of anharmonicity has been tackled using a range of modern methods, including theoretical, experimental, and computational ones, which have given new insights into the microscopic mechanisms of anharmonicity [34-36]. Meanwhile, the concept of phonon "anharmonicity engineering" [35,37] has also been proposed recently to regulate the thermal properties in numerous materials, like thermoelectric materials and thermal barrier coatings [38-40]. For example, in the leading thermoelectric material SnSe, it was recently found that the coupled instability of electronic orbitals and lattice dynamics is the origin of the strong anharmonicity, and thus the ultralow thermal conductivity arises [35,41]. In the typical ionic compound, cesium halides (e.g., CsCl, CsBr, and CsI), a low-temperature anharmonicity was found with anharmonic Cs displacement and short phonon lifetime, compared to the anharmonic behavior commonly occurring at higher temperature [42,43]. The remarkable progress of phonon anharmonicity promotes a deep understanding of thermoelectrics [35,39,41], ferroelectrics [44,45], multiferroics [46-48], and high-$T_c$ superconductors [49-51], etc. Due to the importance of phonon anharmonicity in advanced materials, it is valuable to have a comprehensive review of phonon anharmonicity to help us to understand the fundamental issues in this topic, which is also the motivation of the present article.

1.2 Phonon theory of anharmonicity

Crystal lattice dynamics is based on the concept of phonons, i.e., weakly interacting waves of atomic (or ionic) vibrations and corresponding quasi-particles. A phonon is a quantum mechanical description of a particular type of vibrational motion, in which a lattice uniformly oscillates at the same frequency. In 1932, Tamm first introduced the concept of phonon. Phonon represents an excited state in quantifying the normal modes of vibrations in structures of the interacting particle [52]. Therefore, the lattice dynamics are often referred to as phonon dynamics, and the anharmonicity is often known as phonon anharmonicity in thermodynamics. Here, the theories of phonon anharmonicity will be briefly revisited, and more details can be found in other works [7,53,54].

Due to the smallness of the atomic displacements $\bar{u}$ in crystals compared with interatomic distances $d$, one can pass from a problem of strongly interacting atoms to a problem of weakly interacting phonons. In the leading order in the smallness parameter $\eta = \bar{u}/d$, the crystal lattice dynamics and thermodynamics can be described as ideal phonon gas (harmonic approximation). With the increasing temperature, $\eta$ increases as well. Due to a semi-empirical Lindemann criterion [55] $\eta \approx 0.1$ at the melting point $T=T_m$, higher-order (anharmonic) contributions to the thermodynamic properties are usually small up to the melting temperature [56,57]. However, for some peculiar phonon modes, anharmonicity could be crucially intense and leads to many novel behaviors.

### 1.2.1 Formulation of anharmonicity

One can expand the potential of lattice $V(\{r_j\})$ with $j$ as the site index into the displacement vector $\mu_j = r_j - R_j$, $r_j$ and $R_j$ being the real and equilibrium position:

$$V(\{r_j\}) = V_0 + \frac{1}{2} \sum_{j_1 j_2 \alpha_1 \alpha_2} \Phi_{j_1 j_2}^{\alpha_1 \alpha_2} \mu_{j_1}^{\alpha_1} \mu_{j_2}^{\alpha_2} + \frac{1}{6} \sum_{j_1 j_2 j_3 \alpha_1 \alpha_2 \alpha_3} \Phi_{j_1 j_2 j_3}^{\alpha_1 \alpha_2 \alpha_3} \mu_{j_1}^{\alpha_1} \mu_{j_2}^{\alpha_2} \mu_{j_3}^{\alpha_3} + \cdots,$$

$$\Phi_{j_1 \ldots j_n}^{\alpha_1 \ldots \alpha_n} = \left( \frac{\partial^n V}{\partial \mu_{j_1}^{\alpha_1} \ldots \mu_{j_n}^{\alpha_n}} \right)_{\mu=0}$$

(1)

where $\alpha_n$ is Cartesian indices, and the linear term is absent due to the equilibrium conditions and $\Phi_{j_1 \ldots j_n}^{\alpha_1 \ldots \alpha_n}$ is the $n$th-order interatomic force constant (FC). The quadratic term corresponds to the harmonic approximation, and the higher-order terms correspond to anharmonic ones. The oscillation potentials with displacements between harmonic (second-order) and the third-order anharmonic terms were shown in Figure 1(a), which presents the discrepancy induced by the large atomic displacements.

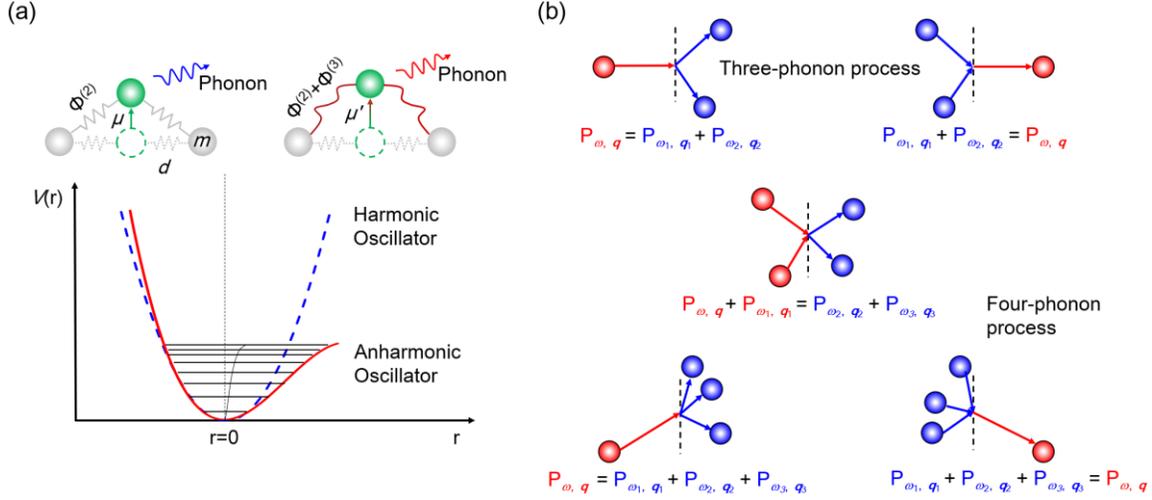

**Figure 1** Comparison of different multiple-phonon scattering processes. (a) Comparison between the harmonic and the anharmonic (3rd-order) phonons. The simple 1D ball-and-spring models represent the harmonic (top left panel) and the anharmonic (top right panel) phonons. The equilibrium state interatomic distance is $d$. The harmonic phonon displaces atom (center, green) by $\mu$ with the linear FC $\Phi^{(2)}$ and the quadratic potential (bottom panel), while the anharmonic one displaces atom by a large $\mu'$ with the nonlinear FC $\Phi^{(2)} + \Phi^{(3)}$ and the quadratic plus cubic potential. (b) Sketches of the three-phonon and four-phonon scattering processes.

In the phonon representation, the total Hamiltonian of the lattice can be expressed as

$$H = V_0 + H_0 + \sum_{n=3}^{\infty} H^{(n)}, H_0 = \sum_{\lambda} \hbar\omega_\lambda \left( b_\lambda^\dagger b_\lambda + \frac{1}{2} \right),$$

$$H^{(n)} = \sum_{\lambda_1 \ldots \lambda_n} \frac{\Phi_{\lambda_1 \ldots \lambda_n}^{(n)}}{n!} A_{\lambda_1} \ldots A_{\lambda_n}, A_{\lambda_i} = b_{\lambda_i} + b_{-\lambda_i}^\dagger \ (i=1 \text{ to } n) \quad (2)$$

where $H_0$ is the Hamiltonian of the ideal phonon gas (harmonic approximation). $\hbar$ is the Plank constant. $\omega_\lambda$ is the phonon energy with the quantum index $\lambda$ [$\lambda \equiv q\xi$, $q$ is the wavevector running the Brillouin zone (BZ), and $\xi$ is the phonon band index]. $b_\lambda$ and $b_\lambda^\dagger$ are the annihilation and creation phonon operators. The multi-phonon scattering matrix elements $\Phi_{\lambda_1 \ldots \lambda_n}^{(n)}$ are the $n$th derivatives of the potential energy with respect to atomic displacement (i.e., $n$th-order force constant). The matrix elements describe the processes of phonon-phonon interaction, such as a merging of two phonons into one, or vice versa, a decay of a phonon into two ($n = 3$), scattering of two phonons into two new states ($n = 4$), etc. [Figure 1(b)].

1.2.2 Phonon self-energy

In classical mechanics, for a generic nonlinear system, the oscillation frequencies mainly depend on the oscillation amplitudes [58]. Therefore, the anharmonicity that leads to the temperature-dependent phonon spectra mostly results from the growth of average oscillation amplitudes with increasing temperature. In quantum mechanics, the same effect can be described in terms of the phonon self-energy [59,60] due to the phonon-phonon interactions. Phonon self-energy is a complex function, which is defined as $\Sigma(\omega) = \Delta(\omega) + i\Gamma(\omega)$. The real part $\Delta(\omega)$, is associated with the phonon frequency shift due to scattering by other phonons, and thus is responsible for the temperature dependence of the phonon frequency. The imaginary part $\Gamma(\omega)$, describing the probability of phonon decay, represents the inverse of the phonon lifetime. The real part of the self-energy can be expressed as

$$\Delta\omega_\lambda = \Delta_\lambda^{(qh)} + \Delta_\lambda^{(3)} + \Delta_\lambda^{(4)} \qquad (3)$$

where $\Delta_\lambda^{(qh)} = -\gamma_\lambda \Delta V(T)/V$ is the quasi-harmonic contribution due to the temperature dependence of the volume $V(T)$, $\gamma$ is the Grüneisen parameter, $\Delta_\lambda^{(3)}$ and $\Delta_\lambda^{(4)}$ are the contributions of the three and four-phonon scattering processes, correspondingly [24,61]:

$$\Delta_{\xi k}^{(3)} = -\frac{1}{2\hbar^2} \varphi \sum_{\eta \varsigma q} \left| \Phi_{\xi k; \eta q; \varsigma, -k-q}^{(3)} \right|^2$$
$$\times \left( \frac{1 + N_{\eta q} + N_{\varsigma, k+q}}{\omega_{\eta q} + \omega_{\varsigma, k+q} + \omega_{\xi k}} + \frac{1 + N_{\eta q} + N_{\varsigma, k+q}}{\omega_{\varsigma, k+q} + \omega_{\eta q} - \omega_{\xi k}} \right. \qquad (4)$$
$$\left. + \frac{N_{\eta q} - N_{\varsigma, k+q}}{\omega_{\varsigma, k+q} - \omega_{\eta q} - \omega_{\xi k}} - \frac{N_{\eta q} - N_{\varsigma, k+q}}{\omega_{\eta q} - \omega_{\varsigma, k+q} - \omega_{\xi k}} \right)$$

$$\Delta_{\xi k}^{(4)} = \frac{1}{2\hbar} \sum_{\eta q} \Phi_{\xi k; \varsigma, -k; \eta q; \eta, -q}^{(4)} \left( 1 + 2N_{\eta q} \right) \qquad (5)$$

where $\varphi$ is the principal value symbol, $N_\lambda = (\exp(\hbar\omega_\lambda/k_B T) - 1)^{-1}$ the Bose occupation, $\hbar$ the Planck constant, $\xi$, $\zeta$, $\eta$ and $\bm{k}$, $\bm{q}$ are the polarization indexes and wavevectors components in the quantum index $\lambda$. At high temperature $T \gtrsim \theta_D$ ($\theta_D$ is Debye temperature), $N_\lambda \simeq k_B T/\hbar\omega_\lambda$, and all three contributions in Eq. (4) are linear in temperature. Figure 2(a) shows the real part of the self-energy of the transverse acoustic (TA) phonon mode at X-point [$q$ = (0, 0, 0.8)] in germanium [62]. It can be seen that the calculated quasi-harmonic, third-order, and fourth-order contributions all show a linear relation with increasing temperature, especially above the Debye temperature ($\theta_D$= 371 K [63]). The quasi-harmonic approximation gives a positive contribution to the energy shift ($\Delta^{(0)}$), while the third-order ($\Delta^{(3)}$) and fourth-order ($\Delta^{(4)}$) results give the negative contribution to the energy shift. The all three contributions jointly drive a negative linear related total energy shift of TA mode with increasing temperature, which indicates the significant role of anharmonicity on the lattice dynamics in germanium.

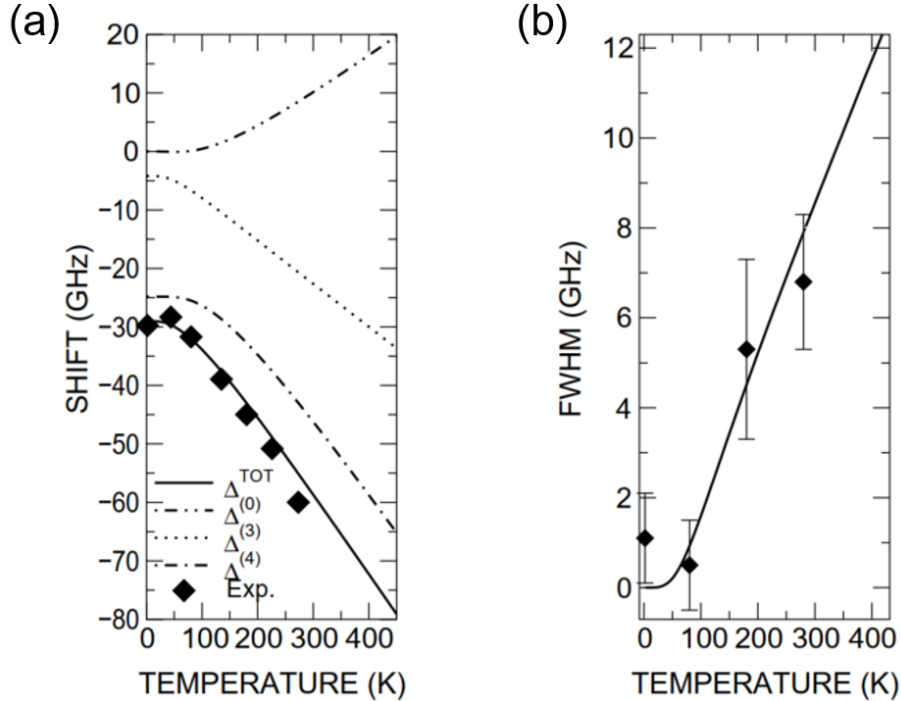

**Figure 2** Real (a) and imaginary part (b) of the self-energy of the TA mode at X-point [$q$ = (0, 0, 0.8)] as a function of temperature in germanium. Diamonds denote spin-echo measurements; solid, double-dot-dashed, dotted, and dot-dashed lines denote the total, first-principles calculated quasi-harmonic, third-order, and fourth-order contributions, respectively [62].

When considering the anharmonicity, the phonons cannot be regarded as stable quasi-particles; e.g., due to the phonon-phonon interactions, they can decay into other phonons.

As a result, the phonon damping (i.e., the inverse phonon lifetime) arises. In the lowest-order perturbation theory, the damping is equal to

$$\Gamma_{\xi k} = \frac{\pi}{2\hbar^2} \sum_{\eta\varsigma q} \left|\Phi^{(3)}_{\xi k;\eta q;\varsigma,-k-q}\right|^2 \left\{(1+N_{\eta q}+N_{\varsigma,k+q})\right.$$
$$\times \delta(\omega_{\eta q}+\omega_{\varsigma,k+q}-\omega_{\xi k}) + (N_{\eta q}-N_{\varsigma,k+q})$$
$$\left.\times\left[\delta(\omega_{\varsigma,k+q}-\omega_{\eta q}-\omega_{\xi k})-\delta(\omega_{\eta q}-\omega_{\varsigma,k+q}-\omega_{\xi k})\right]\right\} \quad (6)$$

The delta functions correspond to the energy and momentum conservation laws for the decay processes. At high temperatures, the damping shows a positive linear relationship with increasing temperature [see Figure 2(b)], i.e., the higher the temperature, the stronger the phonon damping. It is worthwhile mentioning that the fourth-order term contributes only to the frequency shift (real part) but not to the phonon lifetime (imaginary part) [64].

1.3 Scope of this review

In this review, the phonon anharmonicity will be studied and described in a detailed way, including its emergence, driving mechanism, characterization, and some prospects. A general introduction to the background and the theory for phonon anharmonicity are discussed in Section 1. The characterization tools are outlined to study the phonon anharmonicity in both experimental and theoretical methods in Sections 2 and 3. Then, we focus on the origin of the anharmonic behaviors: the intrinsic effects (Section 4) and the evolutions under external stimuli (Section 5) with some examples. Sections 6 introduces a range of materials closely influenced by the anharmonic effects and how the phonon mechanism is incorporated in them. Finally, in Section 7, some remaining challenges, future works, and the prospects of phonon anharmonicity are put forward.

**2. Experimental characterizations of phonon anharmonicity**

After introducing the theory of phonon anharmonicity in the previous section, the characterizations are required to confirm these issues in the real physical systems. As discussed in Section 1, phonon anharmonicity comes from the large atomic displacements in the nonlinear system, which gives high-order contributions to thermodynamic properties. The anharmonic behaviors often manifest as temperature-dependence of mode softening / hardening and phonon damping in dynamics and phonon-phonon scattering processes in kinetics, which lead to the thermal expansion, thermal conductivity, and temperature-dependence of elastic moduli in thermodynamics. There are varieties of methods to study phonon anharmonicity via characterizing these behaviors. It is beyond doubt that the lattice vibration spectrum is one of the most powerful and straightforward approaches to investigate the phonon dynamics, because the inelastic neutron scattering (INS) made it possible for the first time to acquire detailed knowledge of the spectrum of lattice vibrations of a crystal [27]. The temperature-dependence of mode softening/hardening and the phonon damping can thus be mapped out directly, which are the intrinsic reflections of phonon anharmonicity based on the self-energy. In this section, the methods applied for the experimental characterization of such events will be outlined and described.

2.1 Inelastic neutron scattering

Phonons are featured on the collective vibration of atoms based on lattice dynamics. Probing the lattice vibrational spectroscopy is a more effective way to investigate phonon anharmonicity. Neutrons have the same order of energy and momentum as lattice vibrations (e.g., thermal neutrons, the typical energy of about 1–500 meV and wavelength around $0.04^{-1}$ nm.), which can exchange a part of their energy and momentum with an excitation throughout the whole BZ in the system when inelastically scattering from the samples. With properties of charge-neutral and spin, neutrons become a crucial technique that is widely used to study heat, mechanics, magnetism, and crystal structure in the condensed matter [65,66].

By measuring the intensity of neutrons scattered from a sample as a function of wavevector transfer and frequency change relative to the incident monochromatic neutron source, one can measure the Spatio-temporal correlations in the material. By measuring the coherent one-phonon INS cross-section in a single crystal, the phonon dispersion throughout the whole BZ can be obtained directly. In this case, the double-differential cross-section can be expressed as [65]

$$\left(\frac{\partial^2 \sigma}{\partial \Omega \partial E_f}\right)_{coh} = \frac{(2\pi)^3}{2V_0} \frac{k_f}{k_i} S(\mathbf{Q}, \omega) \qquad (7)$$

where $\partial\Omega$ is the element of the solid angle, and $V_0$ is the unit-cell volume. The dynamic structure factor, $S(\mathbf{Q}, \omega)$, is defined by

$$S(\mathbf{Q},\omega) = \sum_{\lambda}\sum_{\tau}\frac{1}{\omega_{\lambda}}\left|\sum_{j}\frac{\bar{b}_j}{\sqrt{M_j}}\exp(-W_j)\exp(i\mathbf{Q}\cdot\mathbf{r}_j)(\mathbf{Q}\cdot\mathbf{e}_{j\lambda})\right|^2 \quad (8)$$
$$\times\langle N_{\lambda}+1\rangle\delta(\omega-\omega_{\lambda})\delta(\mathbf{Q}-\mathbf{q}-\boldsymbol{\tau})$$

where $\mathbf{Q} = k_i - k_f = \mathbf{q} + \boldsymbol{\tau}$ and $\hbar\omega_{\lambda}(\pm) = E_i - E_f$ are the scattering vector and energy transfer to the sample. $E_i$ ($k_i$) and $E_f$ ($k_f$) are the incident and the scattered neutron energies (wavevectors). The + (−) sign indicates that the phonon is created (annihilated) in the scattering process, $\boldsymbol{\tau}$ the reciprocal lattice vector, $\bar{b}_j$ the coherent neutron scattering length, $M_j$ the atomic mass in position $j$ of unit-cell, $\mathbf{e}_{j\lambda}$ the polarization vector of normal phonon mode $\lambda$ for atom $j$. $W_j$ is the Debye-Waller factor defined as:

$$W_j = \frac{1}{2}\left\langle\left[\mathbf{Q}\cdot\boldsymbol{\mu}\binom{l}{j}\right]^2\right\rangle \quad (9)$$

where $\boldsymbol{\mu}$ represents the displacement of atom $j$ in unit-cell $l$.

As described above, the phonon dispersion of a single crystal is determined by the coherent one-phonon INS cross-section measurement, which provides the most detailed dynamics information, such as the mode energy (peak position), phonon scattering rate (full width at half-maximum of the peak), and sometimes multiple phonon-particle scatterings (peak asymmetry) [67,68]. Consequently, the anharmonic phonon behavior can be investigated straightforward by tracing the changes in phonon peaks. By performing the incoherent neutron scattering, polycrystalline samples will provide frequency distribution of the phonons (phonon density of states, DOS), which is essential in its own right, especially for thermodynamic studies [5].

There are two typical techniques for INS measurements: the triple-axis neutron spectrometry (TAS) and the time-of-flight (TOF) spectrometry. TAS generally provides the constant-Q or constant-E scan, which has a significant advantage in seeking a relatively small number of phonon energies at points or along the lines of high-symmetry in reciprocal space. TOF can map phonon dispersion, provide the four-dimentional data for single crystals across large volumes in reciprocal space, or give phonon DOS for polycrystalline. The interested readers can find more details of the principles of INS technologies in ref. [5,65,66,69-71].

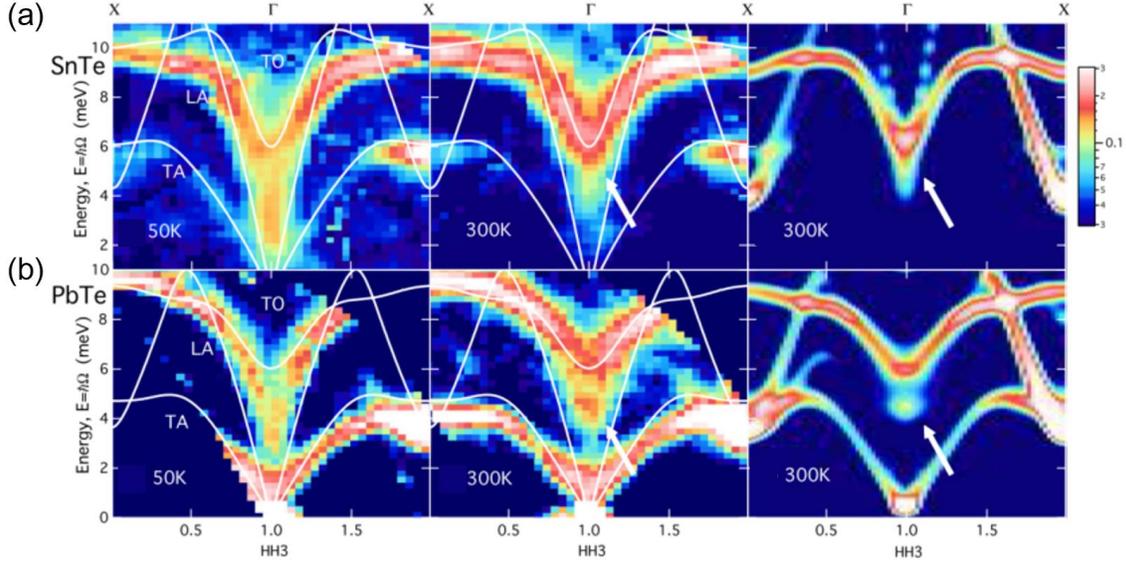

**Figure 3** Anomaly TO modes at the zone center in the $\chi''(\mathbf{Q}, \Omega)$ of SnTe (a) and PbTe (b). The left and the middle panels in (a) and (b) are the INS measurements along [HH3] at 50 and 300K, respectively, while the right panels are the first-principles calculations from phonon self-energy. White lines are the calculated phonon dispersions. Color bars are intensities plotted on a logarithmic-scale [72].

INS has been extensively used to characterize phonon anharmonicity by measuring the phonon dispersion relations and the lifetime of crystals in various sample environments (temperature, pressure, or magnetic). Three recent works are presented here as examples to introduce phonon anharmonicity using both TAS and TOF measurements. The first is to investigate the anharmonic lattice dynamics in SnTe and PbTe by TAS, TOF, and first-principles calculation with anharmonic lattice dynamics [72]. Figure 3 shows that the soft TO modes exist strong broadening and softening with increasing temperature in both materials, indicating the large anharmonicity. In PbTe, the TO modes exhibit a splitting behavior ("new mode" pointed out by arrows), caused by the dispersion nesting between TO and TA. This nesting behavior enables more three-phonon scattering channels in PbTe than SnTe, and amplifies stronger anharmonicity. Such stronger anharmonicity in PbTe mainly results from the sharper resonance in the self-energy. The results reveal that the three-phonon scattering phase space and the combined lattice instability determine the TO ferroelectric soft mode.

The second example is the low thermal conductivity induced by the phonon anharmonicity in thermoelectric material $Ba_8Ga_{16}Ge_{30}$ [73]. The low-frequency optical phonon mode and the acoustic phonons were measured by using the TAS technique, showing an "avoid-crossing" behavior between the TO and LA mode [Figure 4(a)]. This behavior is reflected by the wide phonon linewidths [Figure 4(b)], indicting the strong phonon anharmonicity, and leads to the low thermal conductivity of $Ba_8Ga_{16}Ge_{30}$.

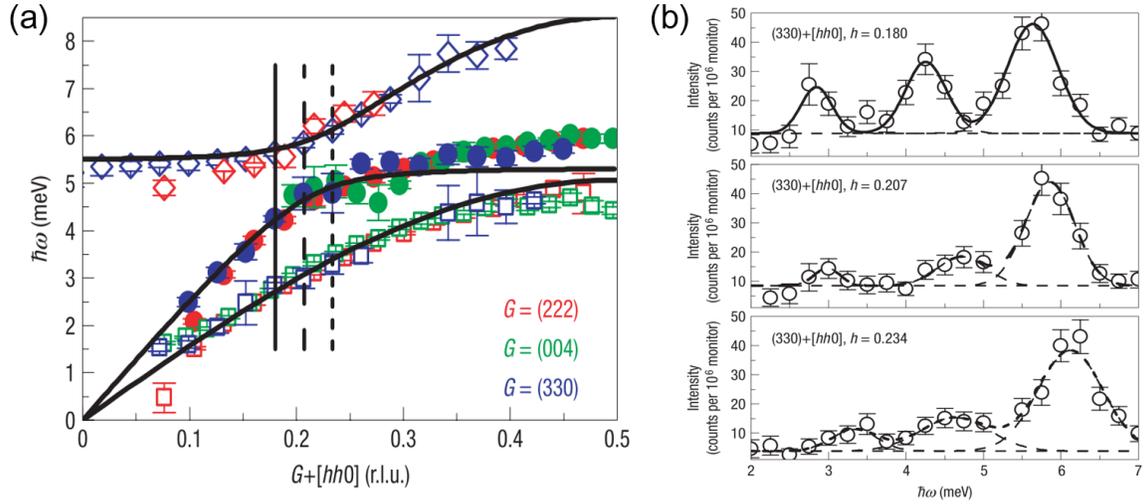

**Figure 4** Measured phonon dispersion of $Ba_8Ga_{16}Ge_{30}$ around the avoid crossing region. (a) The measured phonon dispersion branches along [$hh$0] at G = (222). The error bars represent the fitting deviation. (b) Constant-**Q** scan phonons around the region of the avoid crossing [vertical lines in (a)]. Error bars are from counting statistics [73].

The third one is an example to study phonon anharmonicity by using the TOF technique. In Figure 5(a), It can be found that the measured phonon DOS mostly departs from the quasi-harmonic calculation, especially for the second optical phonon peak, indicating a strong anharmonic system of $FeGe_2$. Figure 5(b) shows the measured $S(\mathbf{Q}, \omega)$ of $FeGe_2$ at 300, 500, and 635K. From the cuts at [400] [Figure 5(c)], the three optical phonons localizing at 18, 21, and 31 meV show clear thermal softening and broadening (shaded region) behaviors with the increasing temperature [Figure 5(d)] [74].

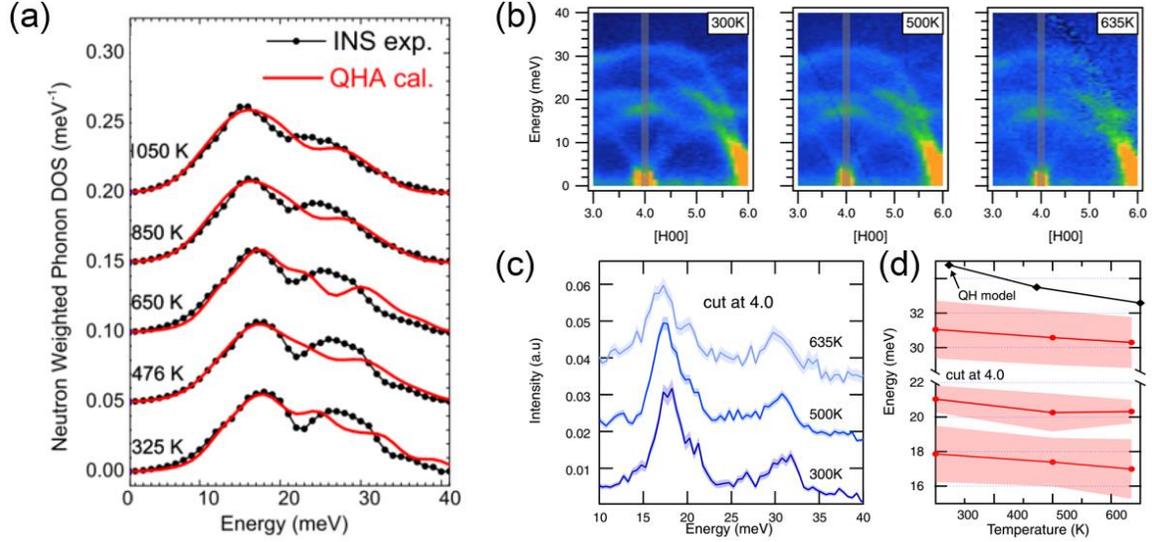

**Figure 5** Phonon DOS and the dispersions of FeGe$_2$. (a) Measured phonon DOS by INS, overlaid by the quasi-harmonic calculation. (b) Measured phonon dispersions along [H00] direction at 300, 500, and 635 K, respectively. The gray line indicates the region of the cut shown in (c). (c) Cuts at 4.0 r.l.u.. Error bars represent the width of the branches. (d) Fitted positions of the optical modes from (c). The red shaded regions represent the linewidth approximated by the peak FWHM. The black line represents the high-energy mode predicted by the quasi-harmonic approximation [74].

2.2 Inelastic X-ray scattering

X-ray diffraction is widely used in crystal structure characterization by focusing on accurate measurements of the positions and intensities of Bragg peaks. In reality, X-ray always acts on the vibrating lattice of crystals, leading to the energy exchange between X-ray and the lattice, i.e., energy loss for a phonon creation and energy gain for a phonon annihilation. Though the high energy of X-ray (KeV) from synchrotron-radiation light source is about six orders of magnitude higher than that of phonons (meV), the technical developments have overcome the challenge in measuring small changes in energies (high resolution, $\Delta E / E < 10^{-6}$), making it possible to determine the complete phonon dispersion. With similar principles of TAS, inelastic X-ray scattering (IXS) also operates the constant-**Q** and constant-*E* scan on single crystals. It is worthwhile mentioning that the IXS allows the micrograms of single crystals samples to be measured for the phonon dispersion. Up to now, this feature makes IXS technology have great advantages in lattice dynamics under high pressure or in the systems in which the big single crystal is a challenge to grow.

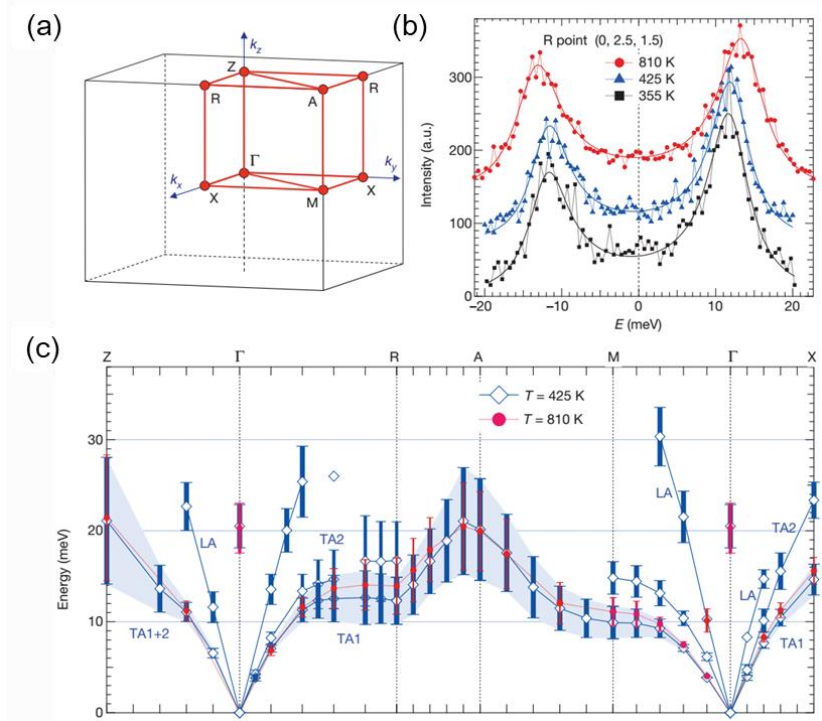

**Figure 6** Phonon dispersions measured with IXS in rutile $VO_2$. (a) Reciprocal space of rutile $VO_2$. (b) IXS scans at R point at 355, 425, and 810 K. Curves are the fitting results by the damped-harmonic-oscillator model. (c) Experimental results at 425 and 810 K. Vertical lines and shading regions indicate the large phonon linewidths [75].

Here, we introduce two recent works to show how IXS is used to investigate phonon anharmonicity. One example is represented on $VO_2$, in which the coherent neutron cross-section of V is very small, and thus IXS is a better choice for investigating the phonon anharmonicity. Figure 6 shows the phonon dispersions of $VO_2$ obtained by IXS [75], in which the phonon anharmonicity is revealed by the prominent broad phonon peaks (short lifetimes) with asymmetric shapes in Figure 6 (b), and all low-energy TA and LA phonons are strongly damped (vertical error bars) in Figure 6(c). In addition, the unusual stiffening of the low-energy TA branches was observed on heating from 425 to 800 K, strongest along Γ–R and Γ–M directions, suggesting that the anharmonic mechanism is responsible for both phonon stiffening and large damping [75].

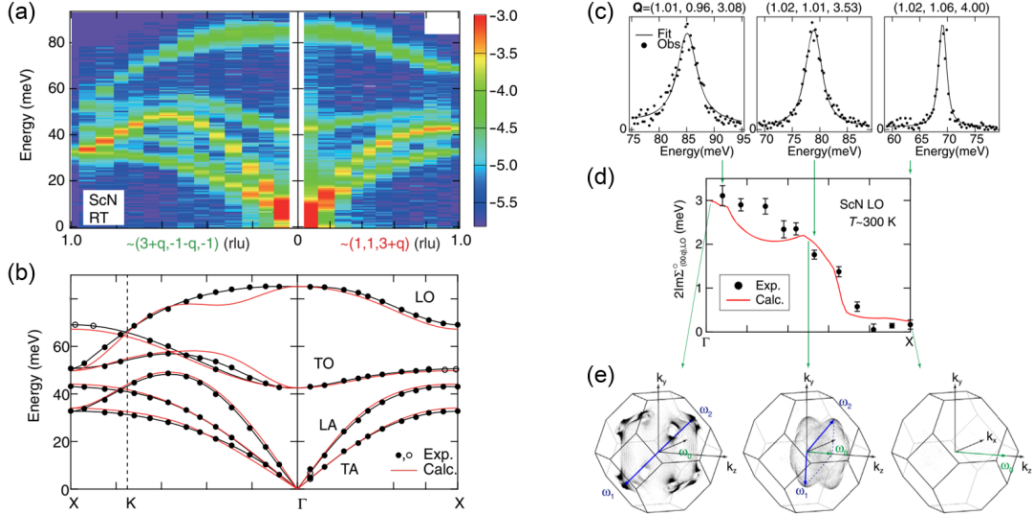

**Figure 7** Phonon properties of epitaxial ScN film [76]. (a) Phonon dispersions measured by IXS along $\mathbf{Q} \sim (3+q, -1, -1)$ (Γ–K–X, left) and $\mathbf{Q} \sim (1, 1, 3+q)$ (Γ–X, right) at room temperature. Color bars represent the intensities on a logarithmic scale. (b) Comparison between the IXS data (black circles) and the first-principles calculation (red lines). Black lines are guides to eyes. (c) Constant-$\mathbf{Q}$ scans of the LO mode by IXS at RT. Lines are the fitting results. (d) Comparison between the extracted (solid black circles) and the calculated linewidths (red line) of LO mode, showing the good agreement. (e) Plots of the calculated decay phonons ($\omega_1$ and $\omega_2$, blue arrows) from the original $(00q)$ phonon ($\omega_0$, green arrow) as a function of $q$ ($q_1$ and $q_2$, black dots) in the reciprocal space of ScN.

The other example using IXS to investigate the phonon anharmonicity is ScN films, which is impossible to carry out by INS due to the small sample size. Figure 7(a) shows the measured phonon dispersions of the epitaxial ScN film measured by IXS along Γ–K–X and Γ–X directions, which is verified by the first-principles calculation [Figure 7(b)] [76]. After fitting the phonon peaks [Figure 7(c)], the LO modes exhibit large linewidths along Γ–X direction. The calculated imaginary part of the self-energy has further confirmed this behavior and is in good agreement with the extracted phonon linewidths (due to the phonon-phonon interaction) [Figure 7(d)]. The details of the phonon scattering were plotted in Figure 7(e), in which one pair of the phonons ($\omega_1$ and $\omega_2$, blue arrows) decays from the original $(00q)$ phonon ($\omega_0$, green arrow) as a function of $\mathbf{q}$. The linewidths of the LO mode also show broadening with the increasing temperature. This decay process is complicated and can be described by both the Ridley (LO → LA / TA +

TO) [77] and Klemens-like (LO → 2LA / TA) [78] channels. Such strong anharmonicity induced by the three-body phonon-phonon interaction may be the origin of the low thermal conductivity in ScN film [76].

2.3 Other experimental methods

In Sections 2.1 and 2.2, we discussed the effective characterization of phonon anharmonicity by INS and IXS, which can measure the phonon spectrum in the whole BZ. This section will discuss other techniques, such as visible light spectroscopy (Raman and Brillouin scattering), Infrared absorption spectroscopy, and electron, which can also probe the lattice dynamics in limited BZ. Due to these sources themselves, the coverage of energy or momentum space is restricted and is only used to measure specific phonon modes or systems. For example, the inelastic electron tunneling spectroscopy (IETS) is sensitive to the vibrations of adsorbed molecules but only from the first few atomic layers on conductive substrates [79]. A latest work reported that the single-defect phonon could be successfully imaged by the newly developed angle-resolved electron energy-loss spectroscopy (EELS); however, resolving phonon dispersions is still a challenge due to the momentum resolutions [80]. Beyond the lattice vibration spectra, some crystalline structural characterizations, such as X-ray and neutron diffraction, X-ray and neutron total scattering, and the extend X-ray absorption function spectroscopy (EXAFS), are also applied to study phonon anharmonicity. In this section, some of the above methods will be briefly introduced, and the interested readers can find more detailed discussions from the related references.

2.3.1 Visible light and Infrared absorption spectroscopy

The magnitude of the wavevector of Infrared (IR) and visible light (Raman and Brillouin scattering) has a typical value of $k = 10^5$ m$^{-1}$, which is tiny compared with the extent of the first BZ ($q = \pi/a \approx 10^9$ m$^{-1}$). Therefore, only the phonons near the zone center can be probed (high-order processes are not involved). The IR photon could be absorbed by the optical mode, showing a sharp peak in the absorption spectrum. This absorption often happens in a polar crystal, in which the transverse optical (TO) phonon mode gives rise to a change of the electrical dipole moment [6,81]. Unlike IR absorption, photons are scattered with a change in energy associated with either the creation of a phonon (energy loss) or annihilation of a phonon (energy gain). There are two types of phonon-related inelastic light-scattering: scattering from acoustic phonons is termed Brillouin scattering, and scattering from optic phonons is referred to as Raman scattering. These two methods are widely used for studying the lattice dynamics, especially for phonon anharmonicity, in various systems such as ferroelectric [82,83], thermoelectric [84,85], and multiferroics [86,87] materials, associated with changes in temperature or pressure.

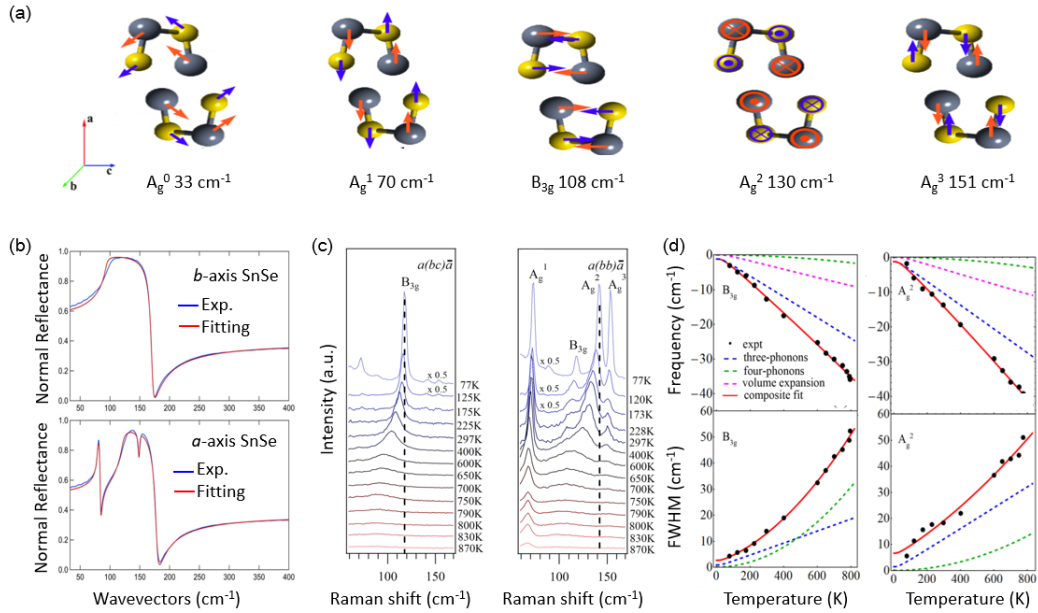

**Figure 8** Phonon anharmonicity investigated by both Raman and IR spectroscopy in SnSe. (a) Raman-active modes in orthorhombic SnSe [90]. (b) Normal reflectance spectra (blue) of $a$- and $b$-axis of SnSe by FTIR at room temperature, overlaid with the Drude-Lorentz fits (red) [88]. (c) Temperature-dependence of the polarized Raman spectra for $B_{3g}$ (108 cm$^{-1}$) mode along $a$-axis (left), and for $A_g^1$ (70 cm$^{-1}$), $A_g^2$ (130 cm$^{-1}$), and $A_g^3$ (151 cm$^{-1}$) modes along $b$-axis (right) [89]. (d) Anharmonic contribution to the phonon softening and linewidth broadening of the $B_3^g$ mode (left) and $A_g^2$ mode (right). The three-phonon, four-phonon, and volume expansion contributions are shown by blue, green, and magenta dashed lines, respectively [89].

Here, we show one example to investigate the phonon anharmonicity in the popular thermoelectric SnSe using IR and Raman. Figure 8(b) shows the normal reflectance spectra along $a$ (lower) and $b$-axis (upper) of SnSe measured by far- transform IR (FTIR) [88], while Figure 8(c) shows the temperature-dependent Raman spectra along *a*- (left) and *b*-axis (right) of SnSe [89]. In comparison, the Raman-active modes [Figure 8(a)] arise at 70 cm$^{-1}$ ($A_g^1$), 108 cm$^{-1}$ ($B_{3g}$), 130 cm$^{-1}$ ($A_g^2$), and 150 cm$^{-1}$ ($A_g^3$) [90], while the IR-active modes arise at 80 cm$^{-1}$ ($B_{1u}^1$), 96 cm$^{-1}$ ($B_{2u}$), 123 cm$^{-1}$ ($B_{1u}^2$), and 150 cm$^{-1}$ ($B_{1u}^3$). It can be seen that the Raman and IR are complementarily used to probe the optical phonon modes due to the different microscopic mechanisms. Moreover, the phonon frequency softening and the linewidths broadening of the $B_{3g}$ and $A_g^2$ are shown in Figure 8(d). According to the Klemens decay model, the softening and broadening contributions are mainly dominated by the three-phonon scattering process [89].

2.3.2 Crystal structure methods

Usually, large lattice vibration amplitude could induce strong phonon anharmonicity. Therefore, investigating the information of a crystal structure, especially the atomic position and atomic distance, could also provide clues to the anharmonicity. However, due to the limitations of lattice dynamics measurements, the crystal structural methods are not as powerful as the vibration spectra. The X-ray and neutron diffraction [91], total scattering [92-94], and the EXAFS [95,96] are often utilized to investigate the anharmonic vibrations, and the main observations related to the phonon anharmonicity are the peak positions, peak widths, as well as peak asymmetry.

Here, we introduce one recent work of ScF$_3$, one of the NTE materials, to show the effectivity of the structural methods to study the phonon anharmonicity by the combination of neutron diffraction, X-ray total scattering, and EXAFS [97]. In Figure 9(a), the lattice constant contracts smoothly with increasing temperature over a large region, showing a prominent NTE behavior. This NTE behavior can be traced to the Sc-F and the Sc-Sc atomic pair distances in the cubic ScF$_3$ from the EXAFS data. The Sc-F bond shows a steep expansion with increasing temperature, while the nearest-neighbor Sc-Sc distance shows a slow shrinking behavior [Figure 9(b)]. The atomic mean square relative displacement (MSRD) can quantitatively investigate the correlated motion between the selected atomic pairs. It can be seen that the MRSD parallel to the bond directions of the Sc-F shows the moderate temperature-dependency [left panel in Figure 9(c)], while the perpendicular one shows the significant-high temperature-dependency [middle panel in Figure 9(c)]. The ratio of the anisotropy (MRSD⊥ / MRSD∥) reaches around 20, which is an extremely large value among the present popular NTE materials [97]. This high anisotropic behavior indicates that the anharmonic transverse thermal vibration of F is the dominant mechanism of the NTE behavior in ScF$_3$. The thermal vibration of F will distort the corner-shared ScF$_6$ octahedrons around the center of the structural frame [dynamic motion in the inset of Figure 9(a)], which shrinks the volume of the system and thus the NTE behavior of ScF$_3$. The phonon anharmonicity of ScF$_3$ was successfully investigated by the crystal methods, and the relationship between phonon anharmonicity and NTE was also studied with INS technique in ScF$_3$ (details can be found in Figure 28 in Section 6).

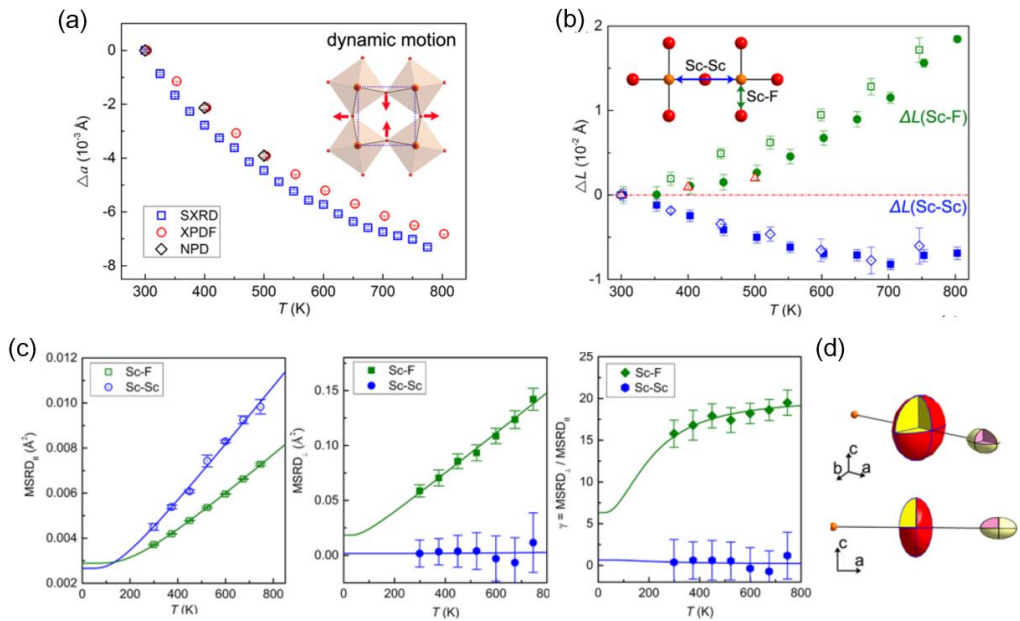

**Figure 9** Temperature-dependent structural properties of ScF$_3$ [97]. (a) Relative lattice constant measured by X-ray diffraction (blue square), X-ray PDF (red circle), and neutron powder diffraction (black diamond). Inset shows the coupled rocking motion of ScF$_6$. Red arrows represent the motion of F. (b) The Sc−F and Sc−Sc (inset) lengths were extracted from the EXAFS (open) and the X-ray PDF (filled). The red triangles represent the Sc−F distance extracted from neutron powder diffraction. (c) Parallel (left), Perpendicular (middle), and the anisotropy atomic MRSDs of the vibrations of Sc-F and Sc-Sc atomic pairs extracted from EXAFS. Curves are fits by the Einstein model. (d) Schematic ellipsoids of the Sc-F (larger) and Sc-Sc (smaller) thermal vibrations.

# 3 Computational characterizations of phonon anharmonicity

As discussed in Section 2, the phonon anharmonicity can be characterized experimentally in varieties of materials by the combination of the vibration spectroscopy and the crystal structure methods. With the development of computational methods and techniques, phonon anharmonicity, especially its microscopic origin, is extensively explored from the first-principles method and ab initio molecular dynamics (AIMD) calculations, such as the DFPT method [98], the DFT-based methods with non-perturbation including the self-consistent phonon (SCP) [99-101] and the self-consistent-field (SCF) theory [1, 102], and the AIMD-based temperature-dependent effective potential (TDEP) method [103-106], etc. In this section, we will not introduce the theory of these methods, and the interested readers can refer to the above literatures. Here, we will give a brief review of typical features to characterize the phonon anharmonicity from simulations, such as non-parabolic frozen phonon potential, large third-order force constants, phonon energy shifting and damping etc. The first-principles method is also powerful to investigate the origin of the phonon anharmonicity, which will be reviewed in Section 4.

## 3.1 Phonon calculations with harmonic approximation

Calculation of harmonic phonon dispersion is usually based on the finite displacement method [107] or the density functional perturbation theory (DFPT) method [108,109]. It is the basis to investigate the lattice dynamics and thermodynamics of weak anharmonic materials. Though the harmonic approximation doesn't work for the strongly anharmonic systems, it can give the sign of the strong anharmonicity of the materials to some extent, such as imaginary phonon frequencies [103,110] and frequency mismatching [103,111]. For example, by using both the PBE and PBE+U functional harmonic calculations [Figures 10(a) and 10(b)], rutile structure $VO_2$ shows the unstable phonon modes, indicating the strong anharmonicity, which was verified from the large phonon linewidth from IXS measurement as shading regions shown in Figure 10(c). With considering the anharmonicity from the TDEP method [103], the unstable phonons are stabilized, and phonon dispersion is in good agreement with the experiment result [Figures 10(c)] [75].

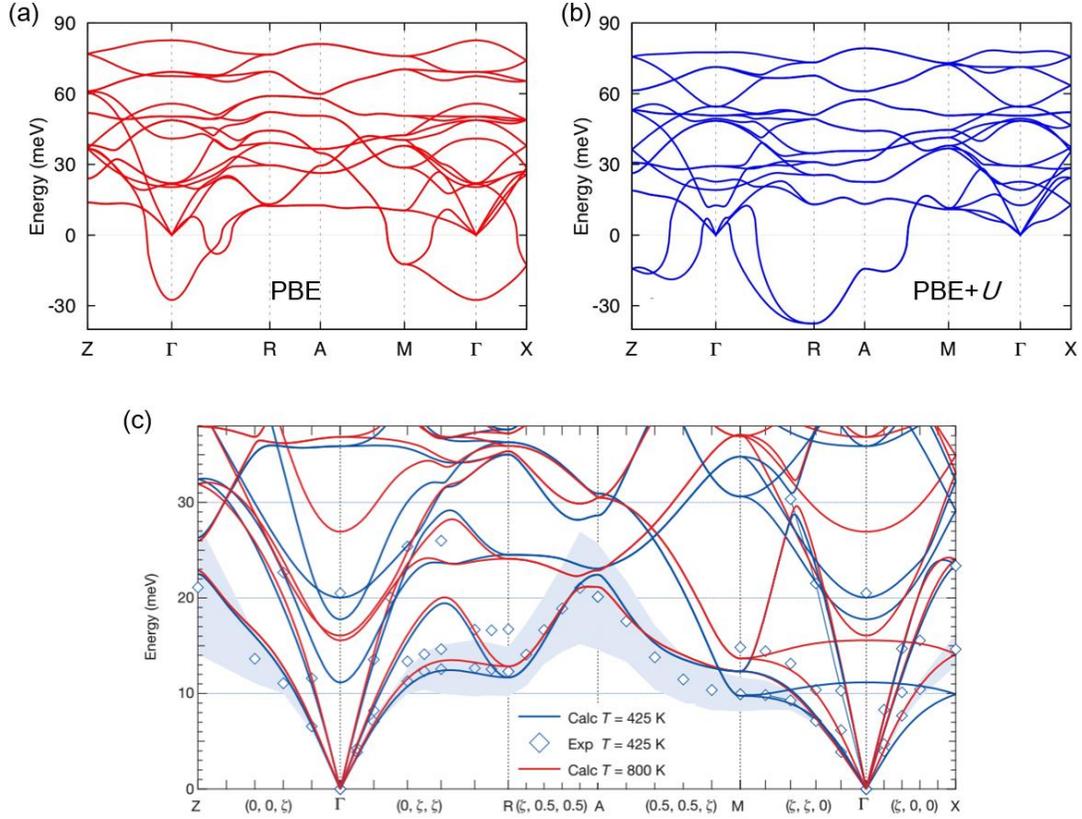

**Figure 10** Comparison between the phonon dispersions calculated from harmonic approximation and TDEP methods with anharmonic effect considered. Results produced by harmonic methods at 0 K: (a) PBE and (b) PBE+U functionals. (c) Phonon dispersions calculated by TDEP methods at 425 (blue lines) and 800 K (red lines). Diamonds and shading regions represent the phonon energies and linewidths at 425 K [75].

3.2 Non-parabolic frozen phonon potential

Harmonic phonons indicate the parabolic frozen phonon potential, which suggests the strong anharmonic system will have non-parabolic phonon potential. Therefore, it is a straightforward approach to explore the anharmonicity by calculating the frozen phonon potential, especially one being interested in a specific phonon mode. The more deviation from the parabolic potential, the stronger the anharmonicity in the systems. As shown in Figures 11(a) and 11(b), when the V atoms vibrate with large displacements, the potential of the mode at R point departs far from the parabolic curve, showing strong anharmonicity of $VO_2$ in rutile phase. However, at low-temperature monoclinic phase, the potential of the same vibration mode follows nearly the parabolic profile, indicating the weak anharmonicity at low temperature phase in VO$_2$ [75]. The similar non-parabolic frozen phonon potential occurs in many advanced materials, such as SnSe [35], ScF$_3$ [112], and Cu$_3$SbSe$_3$ [113] systems (discuss later).

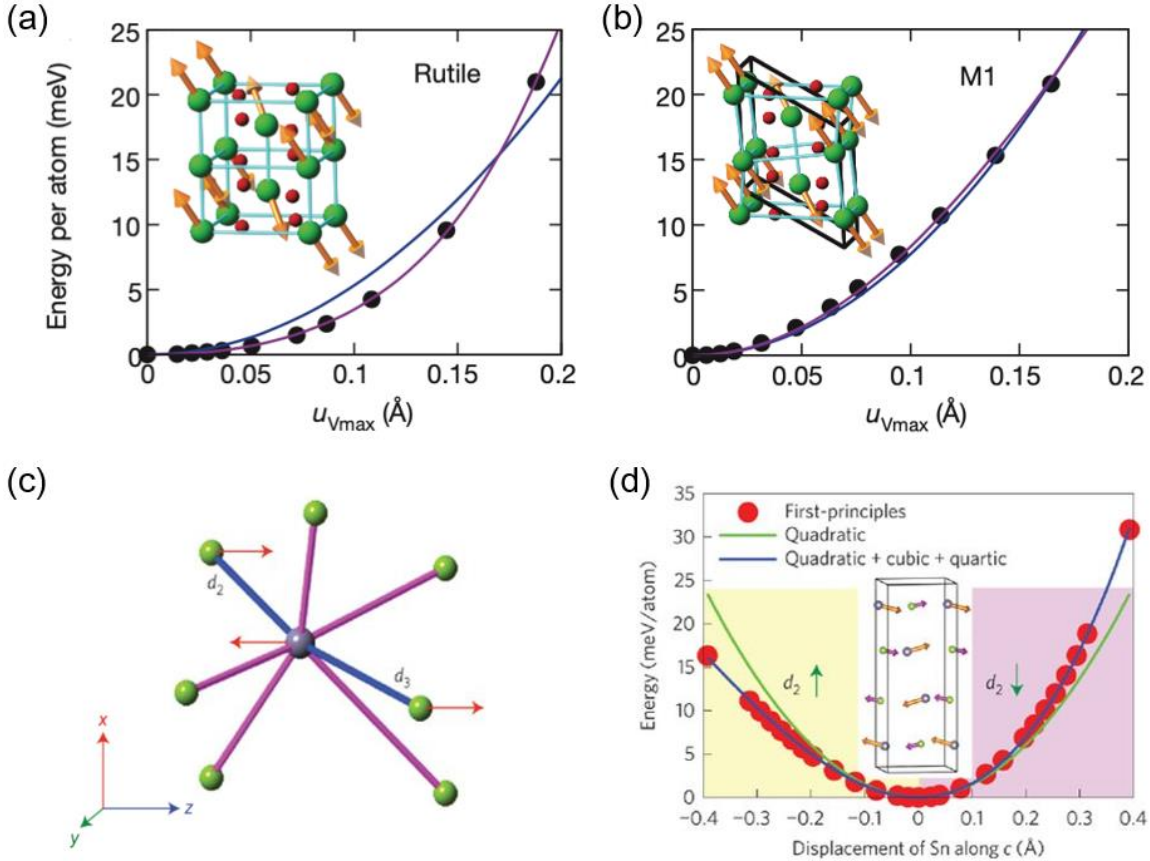

**Figure 11** Frozen phonon potentials to investigate the anharmonicity. (a) and (b) are the frozen phonon potentials of VO$_2$ (a-b) in rutile and monoclinic phases [75]. (c) is the illustration of a strongly anharmonic triplet in SnSe, and (d) is the frozen phonon potential of the TOc mode [35]. Insets are the related vibration modes in these systems.

The frozen phonon potential method is not only a convenient way to evaluate the strength of anharmonicity for a particular mode, but also help to investigate the atomic interactions between specific atoms and explore the origin of the anharmonicity. For example, the Sn atoms in Figure 11(c) vibrate with large displacements, causing the large deviating from the harmonic potential [Figure 11(d)], leading to strong anharmonicity in SnSe. Recently, a large displacement method has been reported to compute anharmonicity without calculating the third-order FCs, in which a larger atomic amplitude corresponding to the given temperature is used to calculate the anharmonic phonon peaks and the dispersion curves [114], which works in cubic W, perovskite $MgSiO_3$, and superconductor $MgB_2$ crystals.

3.3 Third-order interatomic force constant

Based on the interatomic potential expressed in Eq. (1), anharmonicity depends on the high-order terms, which are the most direct quantification of the anharmonicity. Therefore, calculating the third-order force constant is an effective way to characterize the anharmonicity quantitively. It can be seen in Figure 12(a), the Pd atom has square-planar coordination formed by $[Se_2]^{2-}$ and $Se^{2-}$. The larger third-order FCs of $[Se_2]^{2-}$ dimers [Figure 12(b)] reveals that the formation of $[Se_2]^{2-}$ dimers are the origin of the strong anharmonicity in $Pd_2Se_3$ directly [115]. Therefore, it is significant to calculate third-order FC more accurately. There are a few methods to deal with the anharmonic FCs, basically divided into two ways. One is to make a small correction on the harmonic terms by the many-body perturbation theory [67]. This perturbation method is helpful to explore the temperature effects with relatively small anharmonicity [98]. However, due to the challenge of converging higher-order derivatives of the electronic orbitals in DFPT, the application is limited to investigating the systems with large supercells or strong anharmonicity [114].

To overcome the limitations of the perturbation methods, some DFT-based methods have recently developed with non-perturbation. These approximations supplement the conventional treatments, such as the self-consistent ab initio lattice dynamics (SCAILD) [99] and the stochastic self-consistent harmonic approximation (SSCHA) [100], where atoms are displaced in a supercell with the corresponding temperature, and the thermal average of the "exact" potential (or force) is evaluated and does not require a Taylor expansion. In these cases, for example, by the SCP method, the temperature-dependent phonon dispersions can be successfully predicted and show excellent agreement with the experiment [101] [$SrTiO_3$ in Figure 12(c)].

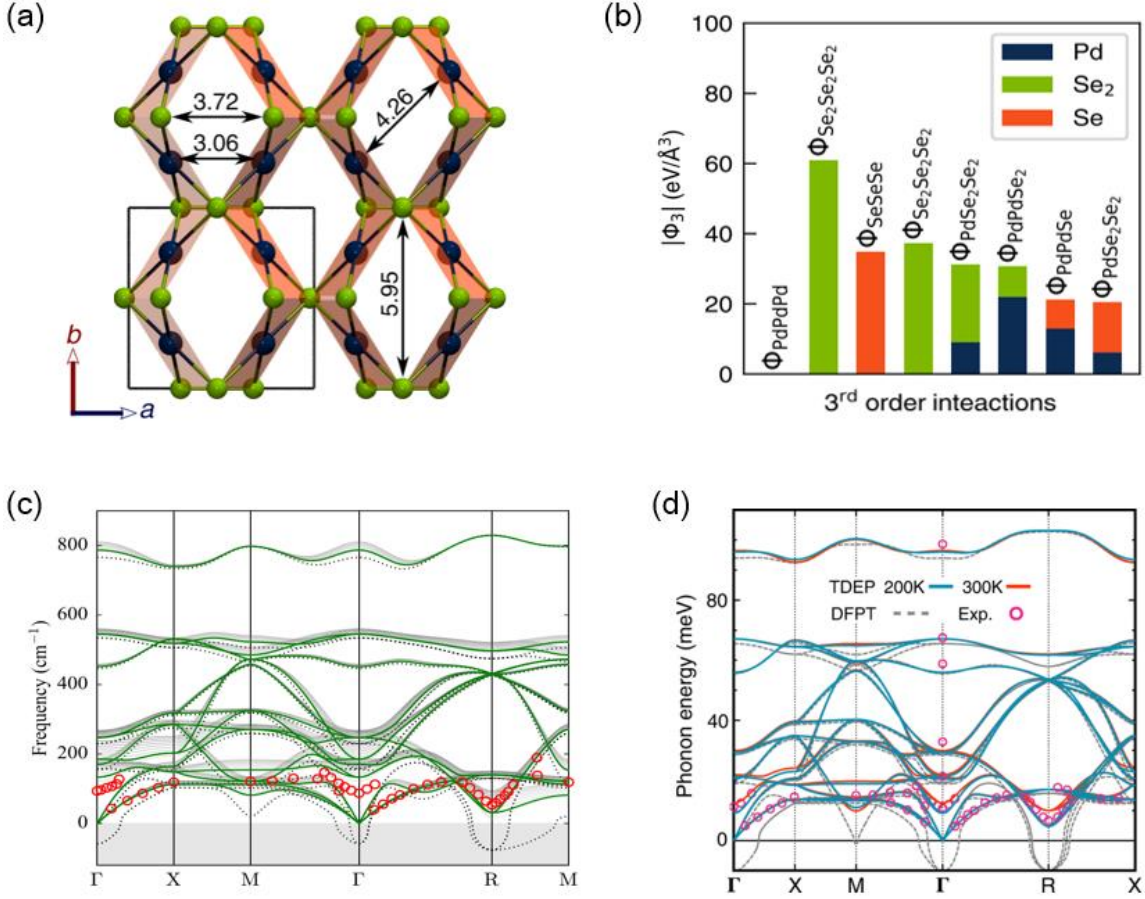

**Figure 12** Different methods to estimate the third-order FCs. (a) Top view of the crystal structure of Pd$_2$Se$_3$. (b) Third-order FCs with different triplets in Pd$_2$Se$_3$ by the DFPT [115]. (c) Phonon dispersions of cubic SrTiO$_3$ by experiments (red circles), harmonic calculations (black dash lines), and the temperature-dependent SCP calculations (solid grey lines, the green lines are highlighted curves at 300 K) [101]. (d) Phonon dispersions of cubic SrTiO$_3$ by experiments (magenta circles), DFPT (grey dash lines), and TDEP (blue and orange solid lines are results at and 200 and 300 K, respectively) [118].

Additionally, the molecular dynamics approach is very powerful in investigating phonon anharmonicity because it naturally includes all effects of anharmonicity in the trajectories while the system is evolving dynamically [103,116,117]. One successful approximation is the TDEP method [103-105,], in which effective third-order FCs are extracted from displace-force data calculated for an atomic configuration sampled by AIMD. Thus, the high-temperature phonon dispersion can be successfully calculated, as shown previous discussed in VO$_2$ [Figure 10(c)] and SrTiO$_3$ [Figure 12(d)] [118]. It shows the TDEP results have good agreement not only with the experiments, but also with the results by SCP method in Figure 12(c).

In most materials, the anharmonicity is mainly dominated by the three-phonon scattering process, i.e., third-order FC successfully estimates the anharmonicity in the system. However, in some systems within (i) strongly anharmonic vibrations, (ii) large acoustic–optical phonon band gaps, and (iii) two-dimensional materials with reflection symmetry, etc., the fourth-order FC should also be taken into consideration [119]. Since 2017, the calculation of fourth-order FC being available from DFT [120], various materials have been investigated successfully, such as BAs [121], graphene [122], and PbTe [123]. It is found that the fourth-order FC plays important roles in these systems, from which the thermodynamics (e.g., the thermal conductivity and the thermal expansion) can be predicted more accurately and reliably.

3.4 Phonon energy shifting and damping

The anharmonic phonon shifting and damping are both determined by the third-order FC according to Eqs. (4) and (6). After obtaining the accurate second and third-order FC, the phonon softening/hardening and the phonon broadening are accessible for further analysis. For example, in silicon, the temperature-dependence of the calculated phonon softening and broadening (linewidth, $1/\tau$) from stochastically initialized TDEP (s-TDEP, including anharmonic and thermal expansion) [124] show good agreement with the experiments (markers), while the results from quasi-harmonic approximation show a large discrepancy with the experiments [Figures 13(a) and 13(b)] [125]. Additionally, the s-TDEP calculated shifts are 65% and broadening 15% different than those of the quasi-harmonic model at 1500 K [Figure 13(c)]. The temperature dependence of shifts and broadenings throughout the BZ in silicon, indicating a strong anharmonic effect in this system [Figure 13(d)].

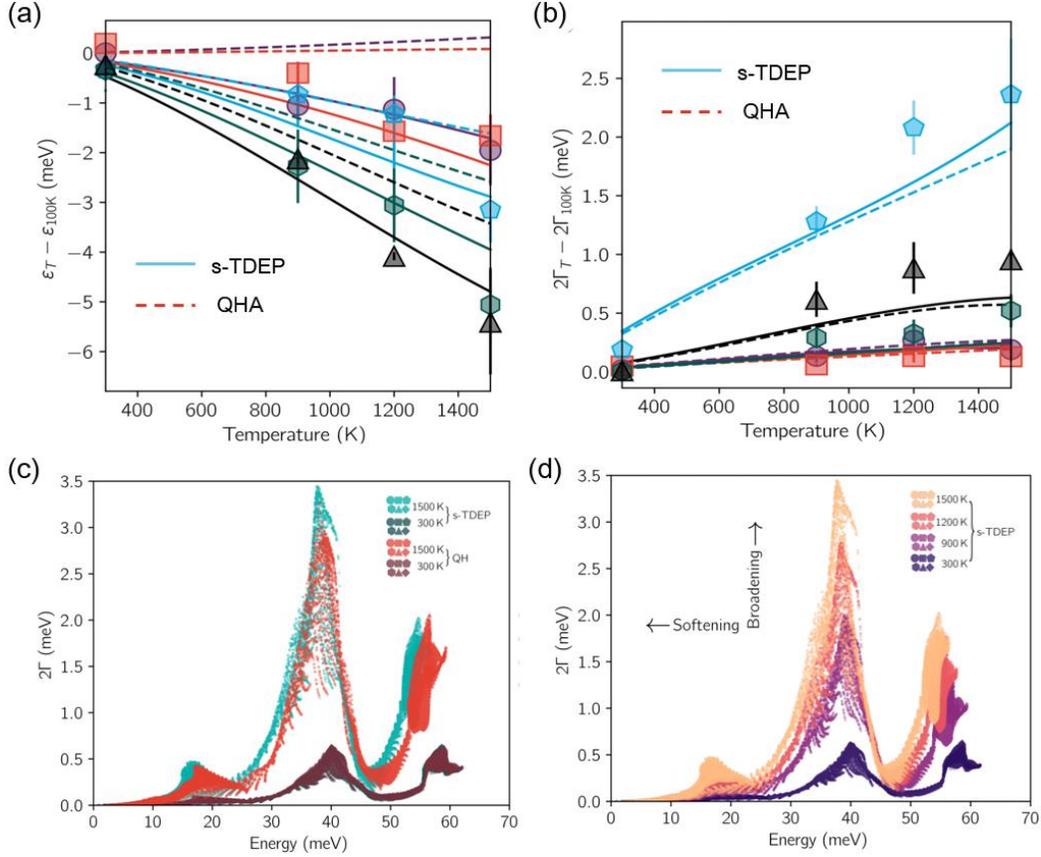

**Figure 13** Phonon anharmonicity of silicon [125]. Temperature-dependence of shifts (a) and broadening (b) of the phonon modes at *q* = (0.75, 0.25, 0.25). Markers, solid, and dash lines represent the experimental, s-TDEP, and quasi-harmonic results, respectively. (c) Comparison between the s-TDEP and quasi-harmonic calculations at 300 and 1500 K. (d) Temperature-dependence of shifts and broadenings throughout the BZ.

As discussed above, depending on the developments of computational methods, various advanced computing methods are accessible for the study of phonon anharmonicity, which provides a microscopic picture of phonon-phonon interactions. By the combination of phonon anharmonicity measurements and calculations, it is possible to obtain a comprehensive investigation of anharmonic lattice dynamics and kinetics, as well as the microscopic origin of the phonon anharmonic interactions.

## 4. Origin of the anharmonic interactions

In previous sections, it shows what phonon anharmonicity is and how it is characterized. However, how the anharmonicity emerges in a real system is complicated to understand.

In recent years, the origin of the anharmonicity has been successfully understood in many systems [35, 126 - 128] by combining phonon measurements and calculations. (as mentioned in Section 2 and Section 3). In this section, we will introduce three typical microscopic mechanisms that lead to the anharmonicity (see Figure 14), including lone pair electrons, resonant bondings, and rattling models.

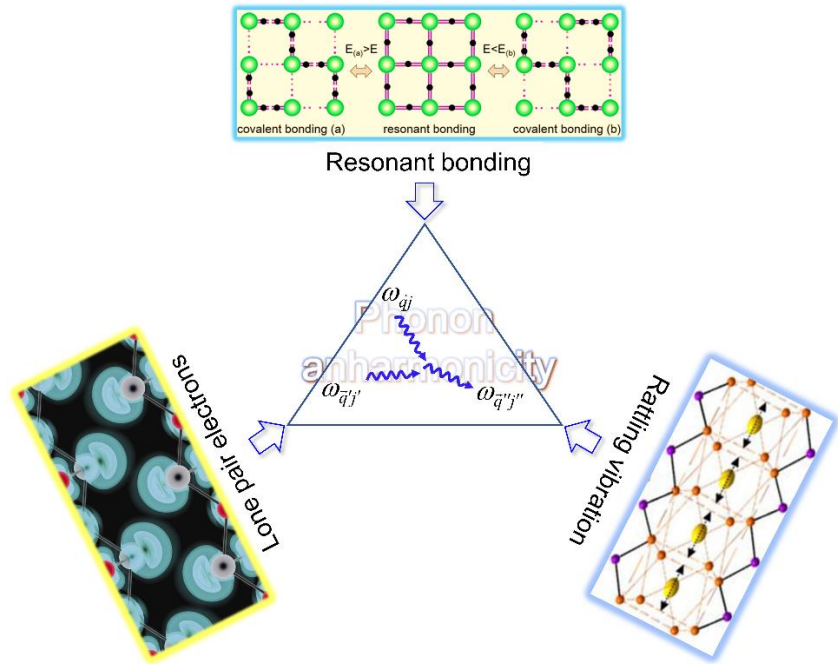

**Figure 14** Three typical mechanisms leading to the intrinsic phonon anharmonicity. The left and the right picture are from ref. [129] and [130], respectively.

4.1 Lone pair electrons

In chemistry, lone pair electrons refer to a pair of outermost valence electrons that are not shared with other atoms. The presence of lone pair electrons (LPEs) will lead to a nonlinear repulsive electrostatic force with the neighboring bonds that lower the lattice symmetry, hinder lattice vibration, and thus induce the bond anharmonicity [131-133]. The formation of LPEs usually requires a strong interaction between the cation $s$ and anion $p$ orbitals and the distortion of the crystal structure [129]. Therefore, for example, for an $ns^2np^m$ valence configured atom, if only the valence electron of $np$ orbitals participate in bonding with the anion, the bond will not share the $ns^2$ electrons and lead the $ns^2$ electron pair to an isolated ("lone pair") state, hence form the LPEs.

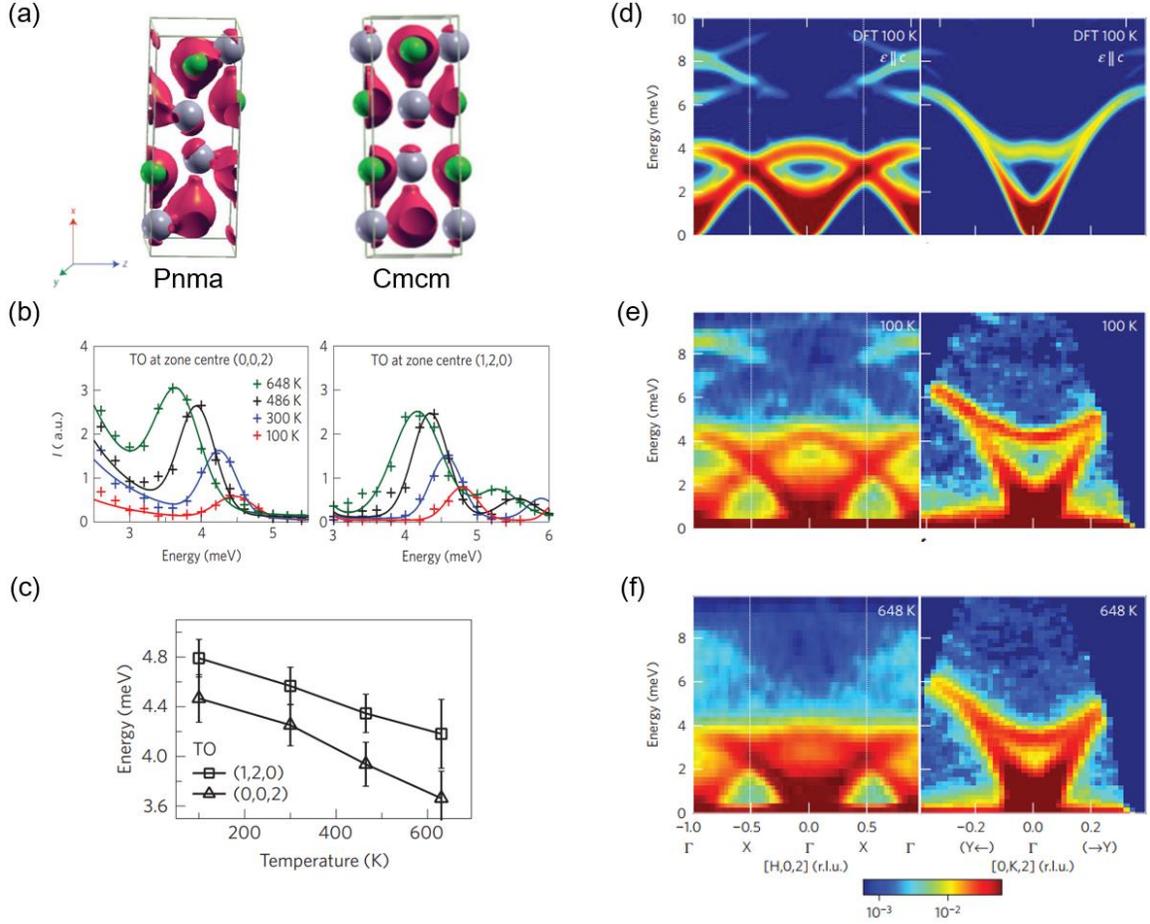

**Figure 15** Strong phonon anharmonicity induced by lone pair electrons in SnSe [35]. (a) Valence electron density (iso-surface at 0.26 eÅ$^{-3}$) for the *Pnma* (left) and *Cmcm* (right) phase structure of SnSe. Sn atoms are in grey and Se atoms in green. The Sn $5s^2$ lone pairs are seen as 'caps' in the space between bilayers. (b) INS spectra for *c* (left) and *b*-polarized (right) TO modes at the zone centers [0,0,2] and [1,2,0], respectively. (c) Temperature-dependence of the TO mode energies. Error bars indicating the phonon linewidths. $S(\mathbf{Q}, \omega)$ along [H02] (left panels) and [0K2] (right panels) directions of SnSe at 100 [calculation in (d) and measurements in (e)] and 648 K (f).

The LPEs induce strong anharmonicity in various materials, such as binary oxides [134-136], binary chalcogenides [35,137-139], ternary oxides [140,141], ternary chalcogenides [132,133,142,143]. SnSe is a typical example to show this issue due to the active $5s^2$ lone pair electrons of Sn atoms, visualized by the 'caps' in the space between bilayers in Figures 15(a) and 15(b), which originate from the hybridization with Se $p_x$-state [35]. Combined with the resonant bonding (discuss later) of *p*-orbitals of Se atoms, the lone pairs induce the off-centering of Sn atoms and the strong Sn-Se bond anharmonicity, especially for the $d_2$ and the $d_3$ bonds in Figure 11(c). These anharmonic bond vibrations cause giant anharmonicity of the TO phonon modes [Figure 11(d)], which are reflected by the temperature-dependent broadening and softening of the TO modes [Figures 15(b) and15(c)] extracted from the INS measurements [Figures 15(d)-(f)]. Consequently, such orbitally driven giant anharmonicity acts as the primary origin of the ultralow thermal conductivity in SnSe.

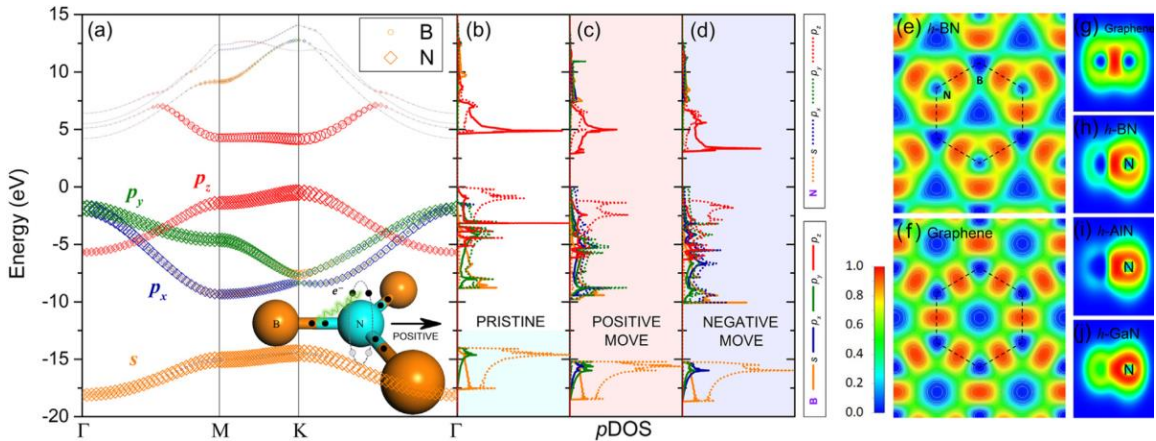

**Figure 16** Lone pair electrons revealed by the first-principles calculations [144]. (a) Orbital projected electronic band structures of $h$-BN. Inset represents the configuration of the lone-pair N-$s$ electrons. Projected electronic DOS of the pristine (b), positively (c), and negatively (d) moved N-atom of $h$-BN. Electron localization function in planar two-dimensional materials (e-j). Top views of $h$-BN (e) and graphene (f), and side views of graphene (g), $h$-BN (h), $h$-AlN (i), and $h$-GaN (j).

This mechanism works for two-dimensional materials as well, in which the stereochemically active LPEs form due to the specific orbital hybridization. For monolayer hexagonal boron nitride ($h$-BN), the $\sigma$ bonds between B and N atoms are jointly contributed by the B-$s$ /$p_x$ /$p_y$ and N-$p_x$ /$p_y$ /$p_z$. The $s^2$ electrons in the $s^2p^3$ valence configuration of N atom do not participate in the bonding, thus forming the lone pair configuration and inducing strong phonon anharmonicity by interacting with the bonding electrons of adjacent B atoms, as revealed by the projected electronic DOS and electron localization function (ELF) (Figure 16) [144].

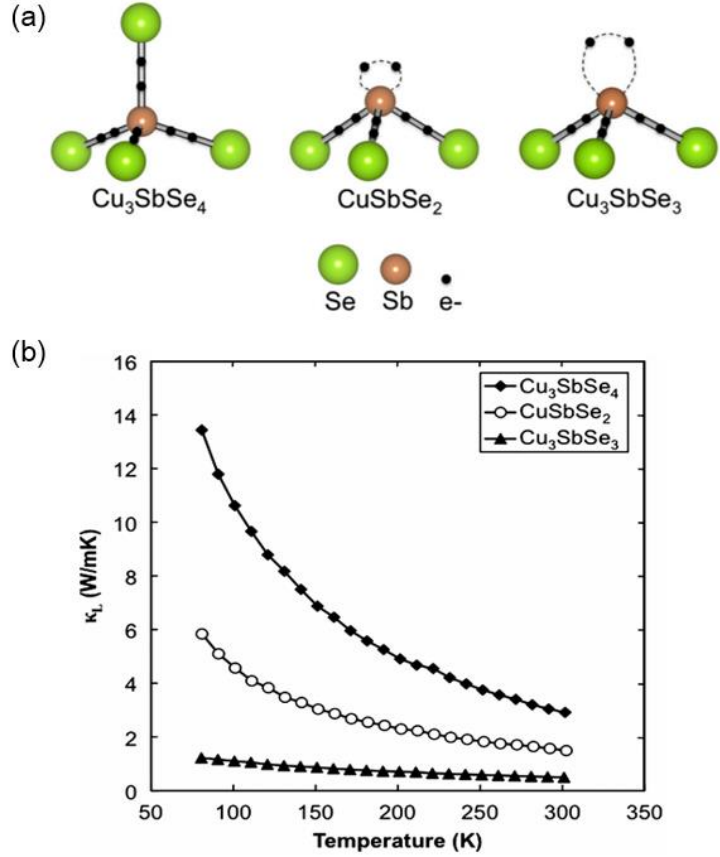

**Figure 17** Low thermal conductivity induced by the lone pair electrons in the Cu-Sb-Se compounds [132]. (a) Schematic plots of the local environment of Sb atoms in $Cu_3SbSe_4$, $Cu_3SbSe_3$, and $CuSbSe_2$. Shaded sticks represent the Sb-Se bonds, and dashed lines represent lone pair $5s$ electron orbital of Sb atoms. (b) Measured thermal conductivities of these three compounds.

LPEs will usually induce the distortion of the bonding structure, which is related to the strength of the repulsive electrostatic force between LPEs and neighboring atoms. The structure distortion mainly depends on the delocalization of LPEs (the distance of LPEs away from the nucleus), which can be visualized by the bond angle in one atom. The Cu-Sb-Se ternary system presents a direct example to study this issue. As shown in Figure 17(a), the bonding angles of Se-Sb-Se are 109.5° for $Cu_3SbSe_4$ without LPEs, while the angles are 95.24° for $CuSbSe_2$, and 99.42° for $Cu_3SbSe_3$ that with LPEs, respectively. In this case, the LPEs are far away from Sb in $Cu_3SbSe_3$ and yet do not form bonding, which leads to the strong bond anharmonicity and hence the lowest thermal conductivity among the three compounds [Figure 17(b)] [132].

4.2 Resonant bonding

Unlike the LPEs, the resonant bonding [145] can also result in lattice anharmonicity even in high symmetry structures, such as the rocksalt and rocksalt-like structures [146-148]. In these cases, a single, half-filled *p*-band forms two bonds to the left and right (more than that is allowed by the 8-N rule, N for the valence), and can be easily understood by the schematic illustrations in Figure 18. For the (001) plane of a hypothetical simple cubic crystal of Sb [147], two valence *p*-electrons of each Sb atom alternately occupy four covalent bonds between the adjacent atoms, as shown on the left and right panels in Figure 18. If the valance bonds change frequently between these two cases and thus reduce the total energy of the system, a superposition of these electronic configurations emerges (middle panel in Figure 18), resulting in the resonance bondings. The remarkable electron delocalization characterized by resonant bonding leads to a significant increase in electronic polarizability.

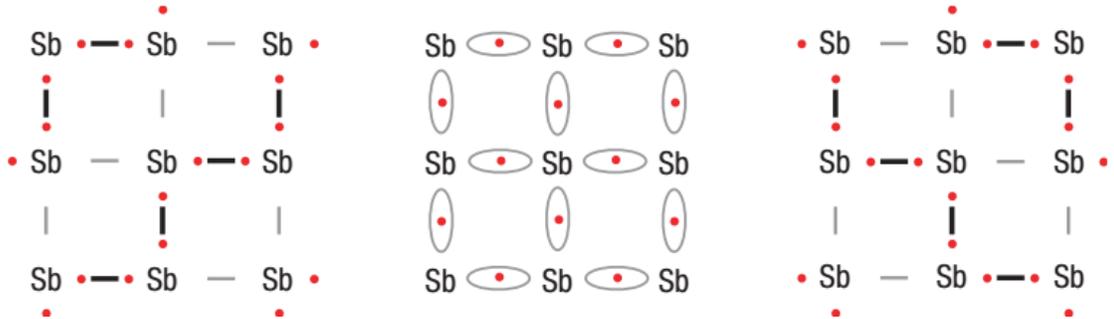

**Figure 18** Schematic plots of origin of resonance bonding for Sb [147]. The same limited cases of bonding states in an undistorted Sb phase are shown in the left and right panels. The resonance bonding phase, formed from the minimizing energy by a hybrid wave function in solids, is shown in the middle panel.

The resonant bonding can commonly induce strong anharmonicity of long-range interatomic FCs in rocksalt structures [149, 150]. Figure 19 shows the trace of the FCs of materials in typical IV-VI, $V_2$-$VI_3$, and V groups [146]. Compared with the NaCl (ionic bonding) and InSb (*sp*-hybridized covalent bonding) shown in Figure 19(b), the long-ranged interactions of the resonant bonding ones (IV-VI, $V_2$-$VI_3$, and V groups) are much larger and non-negligible even for the fourteenth-nearest neighbors.

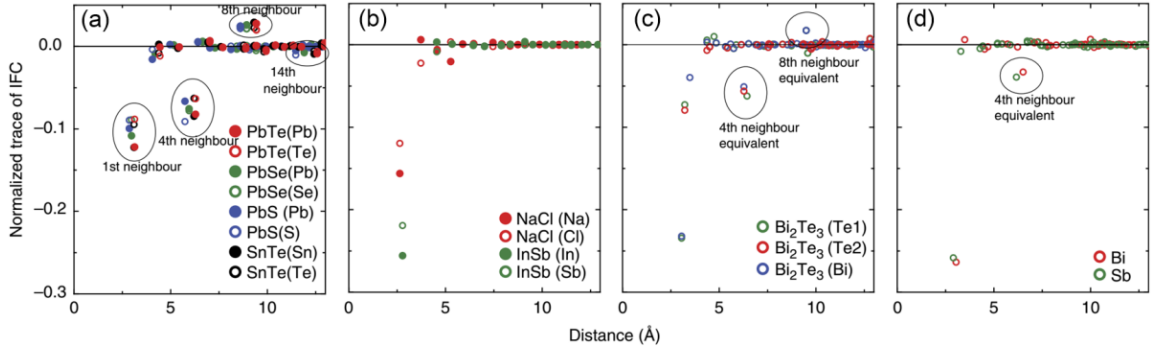

**Figure 19** Normalized traces of the interatomic FCs tensors versus the atomic distances [146]. (a) Lead chalcogenides and SnTe, (b) NaCl and InSb, (c) $Bi_2Te_3$, and (d) Bi and Sb. Circled elements indicate the interactions between the corresponding atom and other atoms.

Figure 20(a) shows the resonant bonding behavior in the rocksalt structure in ferroelectric PbTe. In PbTe, only the *p*-electrons are considered for valence states, and each atom has three valence electrons on average (Pb: 2 and Te: 4). This bonding configuration will form a long-range resonance bond of PbTe, leading to the softening of the TO modes [146,151,152]. It was reported that there is a large coupling between the TO and LA modes due to the softening of the TO modes [153]. Such behavior was later experimentally confirmed by the INS measurement, as shown in Figure 20(b) [151]. It reveals an avoid crossing between LA and TO branches around $q$ = (0, 0, 1/3) due to anharmonic repulsion between these modes, in contrast with the harmonic dispersions (shown as white solid lines) calculated by DFT. Such anharmonic TO mode leads to an intrinsic scattering of LA modes and thereby largely suppresses the thermal conductivity of PbTe, which is significant for thermoelectric applications, shown in Figure 20(c) [154].

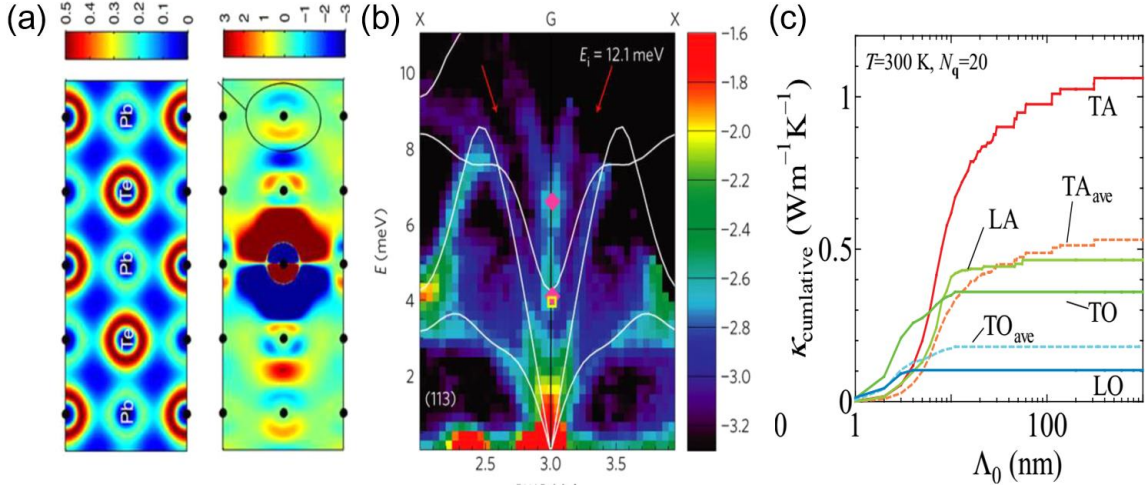

**Figure 20** Strong anharmonicity induced by the resonant bonding that leads to the low thermal conductivity of PbTe. (a) ELF maps of the ground-state of PbTe, revealing the large delocalized electron density (left panel) and long-ranged polarization by the displacement of the center atom (right panel). The circled state represents the fourth-nearest neighbor electronic polarization. Black dots represent the atoms on the (100) plane [146]. (b) Phonon dispersions obtained from INS for PbTe at 300 K, showing the avoided crossing of LA and TO phonon branches. The harmonic dispersions calculated with the DFT are shown as white solid lines [151]. (c) Cumulative thermal conductivity $\kappa_c$ as a function of the mean free path (MFP). TA$_{ave}$ represents the averaged values of the two TA branches. The much smaller value of LA originates from the strong scattering by the anharmonic TO phonons [154].

4.3 Rattling vibrations

The rattling vibration can be described as follows: the guest atoms (or molecules) are weakly bound and somewhat independent of the other atoms in an oversized atomic cage, which vibrate anharmonically with large displacements. Such a rattling behavior was first described by Sievers [155] and was first introduced into thermoelectric materials with cage structures by Slack [156]. Figure 21(a) shows the anharmonic potential induced by the rattling vibration of off-center guest Ba atoms ($6k$ site) in type-I clathrate $\beta$-$Ba_8Ga_{16}Sn_{30}$. The potential of the rattling Ba atoms can be described as $V=Ax^2+Bx^4$, containing a harmonic (quadratic) term and an anharmonic (quartic) term [157,158].

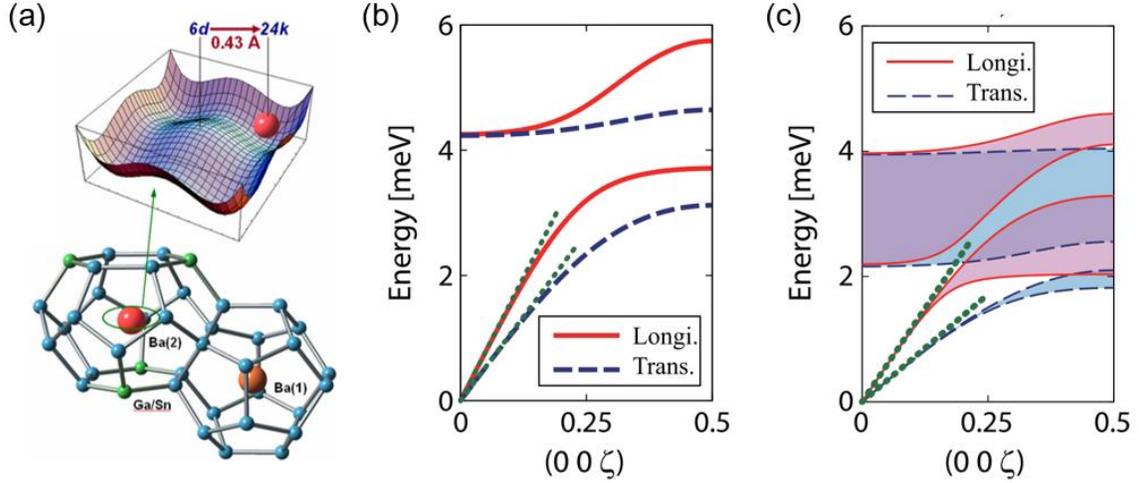

**Figure 21** Phonon anharmonicity in cage structure. (a) Schematic of the cage structure and guest potential well for the Ba (2) in $\beta$-Ba$_8$Ga$_{16}$Sn$_{30}$. The fourfold potential minima (24$k$ site) are offset by 0.43 Å away from the cage center (6$d$ site) of the cage [157]. (b) Phonon dispersions for the on-center Ba$_8$Ga$_{16}$Sn$_{30}$. (c) Phonon dispersions for the off-center $\beta$-Ba$_8$Ga$_{16}$Sn$_{30}$. Solid and dash lines represent the longitudinal and transverse modes, respectively. The shaded regions bounded by the solid and dash lines indicate the broadening induced the off-center vibrations. Dotted lines represent the long-wave limited acoustic branches [163].

In general, one of the main features of the guest atoms inducing rattling vibrational behaviors is that when the available space in the host cages becomes larger than their ionic radii, the restoration forces on them will become weaker. The phonon frequencies will decrease accordingly and behave anharmonically in a more localized region. Therefore, these guest phonon dispersions become low-lying and very flat [159-161]. The rattling guest atoms usually have two types of positions in the cage: on-center and off-center, which depend on the relative size of the cage to the guest atom [162]. As a result, the on-center and the off-center atoms will lead to the difference between their respective phonon dispersions (Figure 21(b) and Figure 4 for the on-center one) [163]. Compared with the on-center one, the phonon linewidth of the off-center one becomes much broader and gapless in the region of the avoid crossing [Figure 21(c)], which is attributed to the random orientation of guest atoms.

Recently, rattling vibration has been widely observed in materials with cage structures (e.g., skutterudite [164] and clathrate [73,157,162]), non-cage structures (e.g., $CsPbI_3$ [165], and $Cu_{12}Sb_4S_{13}$ [166]), and even two-dimensional structures (e.g., $Mg_3Sb_2$ [167]). This behavior can be effectively studied by simulations and experimental techniques, such as X-ray and neutron diffraction [168,169], EXAFS [170], inelastic X-ray and neutron scattering [73,171-173], Raman spectroscopy [161,174], and optical conductivity [175]. Though the rattling phenomenon has been extensively studied, the origin of the low-lying soft modes is still not fully understood, except for the case of lone pair electrons [166,176].

Unlike the rattling motion that is vibrating locally in cage structures, similar states could exist in which the weakly bonded atoms vibrate with much larger thermal atomic displacement parameters (ADPs) in the system, called rattling-like motion [126]. In materials such as filled skutterudites and $Cu_3SbSe_3$ [Figure 22(a)], this vibration will lead to the atomic-level heterogeneity and the mixed part-crystalline-part-liquid structure, and thus the rattling-like thermal damping due to the collective soft-mode vibrations, which is similar to the Boson peak in amorphous materials [177]. Figure 22(b) shows, in $Cu_3SbSe_3$, the large ADPs of a set of Cu atoms (Cu1$z$ atoms vibrate along $z$-direction and Cu2$xz$ atoms vibrate along $x$-$z$ plane) which dominate the rattling-like vibration, presenting the strong anharmonic potentials. Accordingly, such rattling-like mode induces the fluid-like flow of atoms and presents a more thermal damped peak in the phonon density of states compared to the single rattling mode, as shown in Figures 22(c) and 22(d). As a result, the dynamic fluctuations of a set of atoms, which are weakly bonded to the rest of the lattice, can cause the part-crystalline-part-liquid states, leading to strong phonon damping.

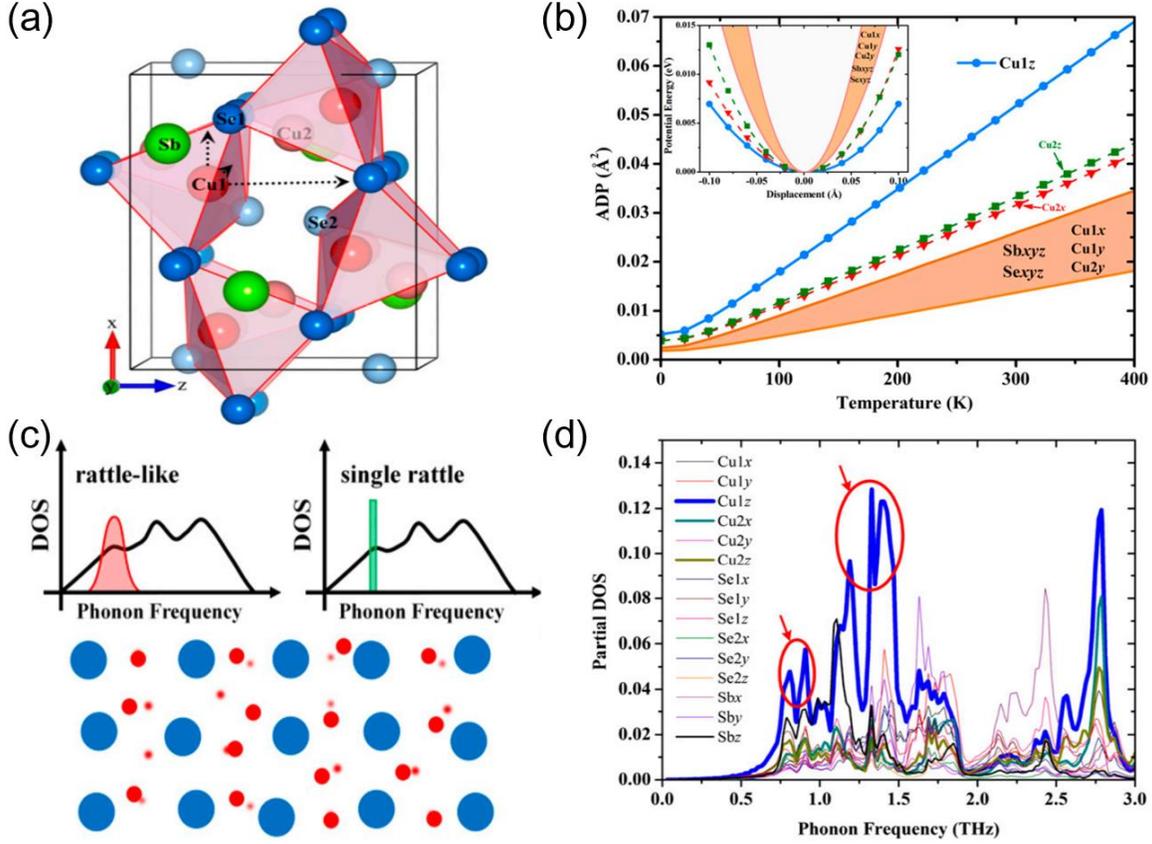

**Figure 22** Ratting-like motion in $Cu_3SbSe_3$ [126]. (a) Crystal structure of $Cu_3SbSe_3$. (b) Calculated ADPs and the corresponding atomic projections for the different atoms. Inset is the calculated potentials for the five nonequivalent atoms. (c) Schematic plots of the rattling-like thermal damping (upper left) in materials with chemical bond hierarchy and single-rattle phonon in filled $CoSb_3$ (upper right). The lower panel represents the part-crystalline part-liquid structure. (d) Partial phonon DOS of $Cu_3SbSe_3$. *x*, *y*, and *z* indicate the vibration directions.

## 5. Effect of external stimuli

In Section 4, we discussed the intrinsic origin of phonon anharmonicity and introduced how the electrons, atoms, bondings, and phase structures account for the anharmonic behaviors. However, phonons are very sensitive to the external environment, such as temperature [36,74,124], pressure [141,178], electric fields [179,180], and magnetic fields [181,182], which may dramatically influence the lattice dynamics and cause various novel phenomena in solids. Herein, the temperature and pressure-induced phonon dispersive behaviors will be shortly introduced. The interested readers can find more details from the cited works above.

5.1 Temperature effect

Theoretically, the temperature-dependent effect is the primary stimulus that leads to the anharmonicity introduced in phonon systems [6]. At elevated temperatures, the higher-order (anharmonic) terms in Eq. (1) cannot be negligible, and the anharmonic effects will become more and more prominent with increasing temperatures. The temperature-effects, such as the phonon frequency shifting and damping, were discussed in passing in various materials above. Here, we will discuss another interesting anharmonic phenomenon related to the temperature, which is known as the liquid-like thermal vibration in the superionic compounds, such as $Cu_{2-\delta}X$ ($X$ = Se /S) [183,184], $MCrX_2$ ($M$ = Cu /Ag, $X$ = Se /S) [185-188] and $AgCuX$ ($X$ = S /Se /Te) [189,190]. In these materials, the lattice exhibits crystalline order, while the atoms follow the long-range diffusion [191]. Thus, the thermal conductivity is strongly reduced in a solid-state by a remarkable strong anharmonic effect — the breakdown of acoustic phonon transport in a host lattice with a liquid-like component, based on the notion that liquids do not propagate shear waves [192].

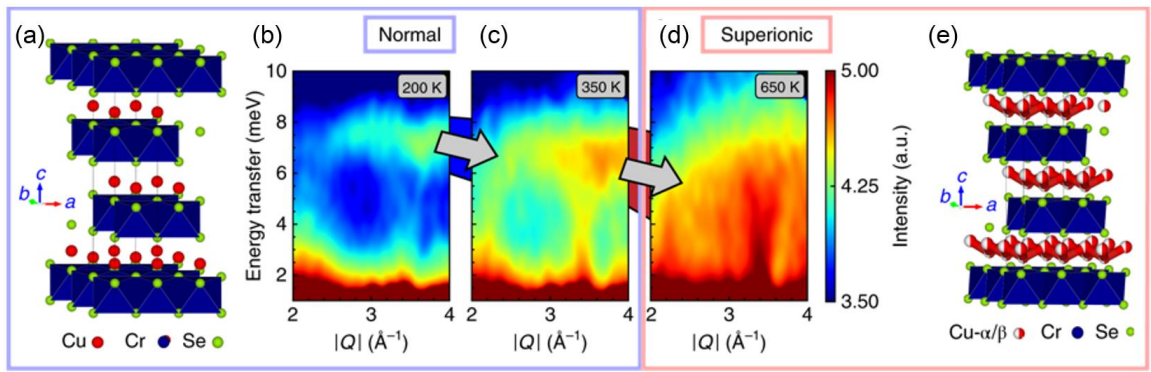

**Figure 23** Anomalous atomic dynamics in $CuCrSe_2$ [197]. Crystal structures of room- (a) and high-temperature (e) phase of $CuCrSe_2$. The half white-half red spheres represent the occupancy of Cu atoms on both $\alpha$ and $\beta$ sites in the high-temperature phase. (b)-(d) $S(\mathbf{Q}, E)$ measured by INS at 200 (b), 350 (c), and 650 K (d). (b)-(d) show the dispersive acoustic phonons remain visible in the superionic phase.

In this case, the acoustic phonons (both the longitudinal and the transverse phonons) are significantly damped as temperature rises in the ordered phase. Interestingly, above the order-disorder transition temperature, the transverse acoustic phonons are dramatically suppressed (melting) through the ultrafast dynamical disorder. Such behavior may be explained beyond the rigorous definition of phonon, which is a collective excitation in a well-defined lattice. Simultaneously, the LA mode is strongly scattered but survives and is thus responsible for the intrinsically ultralow lattice thermal conductivity [193, 194]. Despite such context, it has still been argued that the strong anharmonic low-lying rattling phonons could be the source of thermal transport suppression [171,195,196]. For example, in a recent work of $CuCrSe_2$, there exists an order-disorder transition around 363 K [197] (Figure 23). At low temperature [Figure 23(a)], the Cu ions occupy only the $\alpha$ sites (R3m symmetry); warming across $T_{od}$ ~ 363 K [Figure 23(e)], the Cu ions across $\alpha$ and $\beta$ sites with equal occupancies; above $T_{od}$, the Cu ions undergo 'superionic' quasi-2D diffusion in their occupied sublattice ($R\bar{3}m$ symmetry) [198]. By tracking the evolution of the $CuCrSe_2$ lattice dynamics from 10 to 650 K [Figure 23(b)-(d)], it can be seen that the low-energy phonon band (~ 8 meV) softens and broadens on heating and eventually spread over most of the low-energy transfer range above $T_{od}$. This behavior shows a strong anharmonic effect with increasing temperature; meanwhile, the intensities of Bragg peaks and acoustic phonons (both LA and TA) remaining visible.

Such behavior can be further understood from Figure 24, in which the low-energy peak of phonon DOS noticeably softens and broadens with increasing temperature [Figure 24(a)]. In Figures 24(b) and 24(c), the projected partial phonon DOS reveals that the 8 meV peak arises from in-plane phonons dominated by the vibrations of Cu atoms, which indicates the Cu atoms undergo large-displacement quasi-two-dimensional (quasi-2D) vibrations in this energy region in the ordered phase. Subsequently, the vibration of Cu atoms becomes delocalizing and diffusive upon heating into the superionic phase, leading to an overdamped peak (large anharmonic oscillations). This strong anharmonic vibration of Cu atoms is well revealed by the AIMD trajectories [Figures 24(d) and 24(e)]. It is found that the Cu atoms localize around the $\alpha$ site at 300 K (ordered), while delocalize across $\alpha$ and $\beta$ (z = 1) sites at 500 K (disorder). This evidence reveals that the temperature-dependent anharmonic TA modes remain well defined through the superionic transition. In contrast, the specific phonon quasi-particles break down due to lattice anharmonicity and disorder.

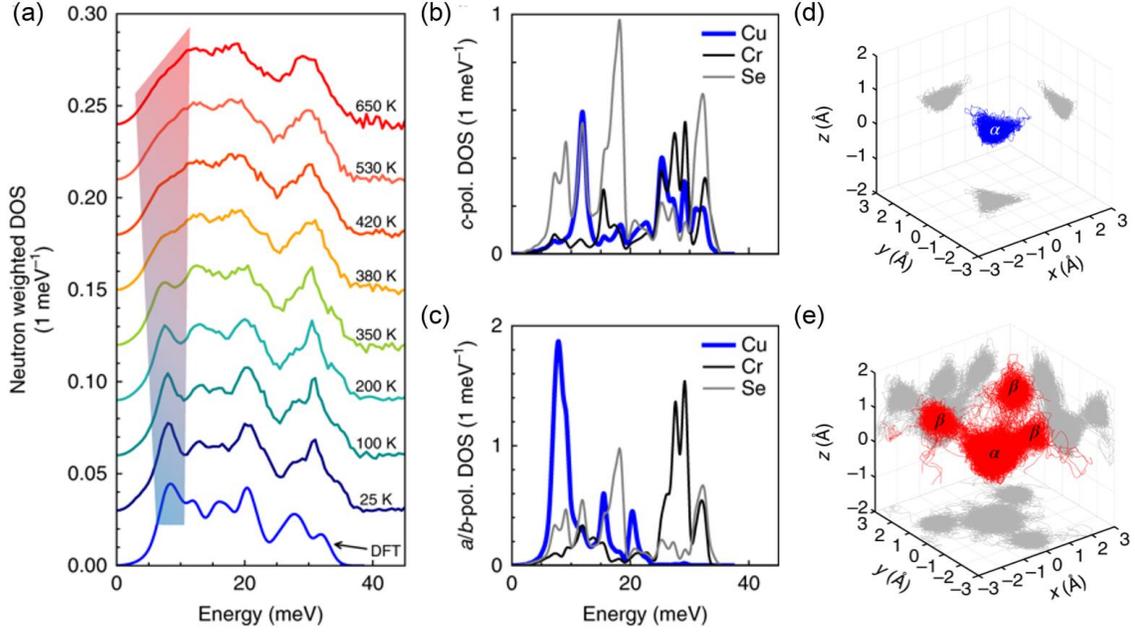

**Figure 24** Temperature-dependent anharmonic evolutions of Cu-dominated phonon modes [197]. (a) Neutron-weighted DOS with increasing temperatures. The trace labeled by shading is the DFT results renormalized by neutron weighting and convolution with the instrument energy resolution. (b) and (c) are the out-of-plane and in-plane DOS, respectively. The AIMD trajectories for the Cu atoms show the localization of Cu atoms around $\alpha$ site (z = 0) at 300 K (d), compared to the delocalization of Cu atoms across $\alpha$ and $\beta$ (z = 1) sites at 500 K (e).

5.2 Pressure effect

It is worthwhile mentioning that the temperature effect possesses two components: changing the lattice volume and increasing the vibrational amplitude of atoms. The former is known as the implicit anharmonicity, i.e., the quasi-harmonic behavior in lattice dynamics, while the latter is known as the explicit anharmonicity dominated by phonon-phonon interactions [178,199]. Pressures affect the equilibrium spacings between nuclei and distort the electronic cloud, and thus the volume of the lattice is changed. Although the phonon frequencies also shift versus pressure at a fixed temperature, this mechanism ignores the vibrational amplitude of atoms, which can only be regarded as the quasi-harmonic behavior. However, it is beneficial to decouple the pure anharmonic and quasi-harmonic effects by comparing the temperature and the pressure measurements (Section 5.3). Therefore, it is necessary to have a brief introduction to the pressure-induced phonon frequency shifts.

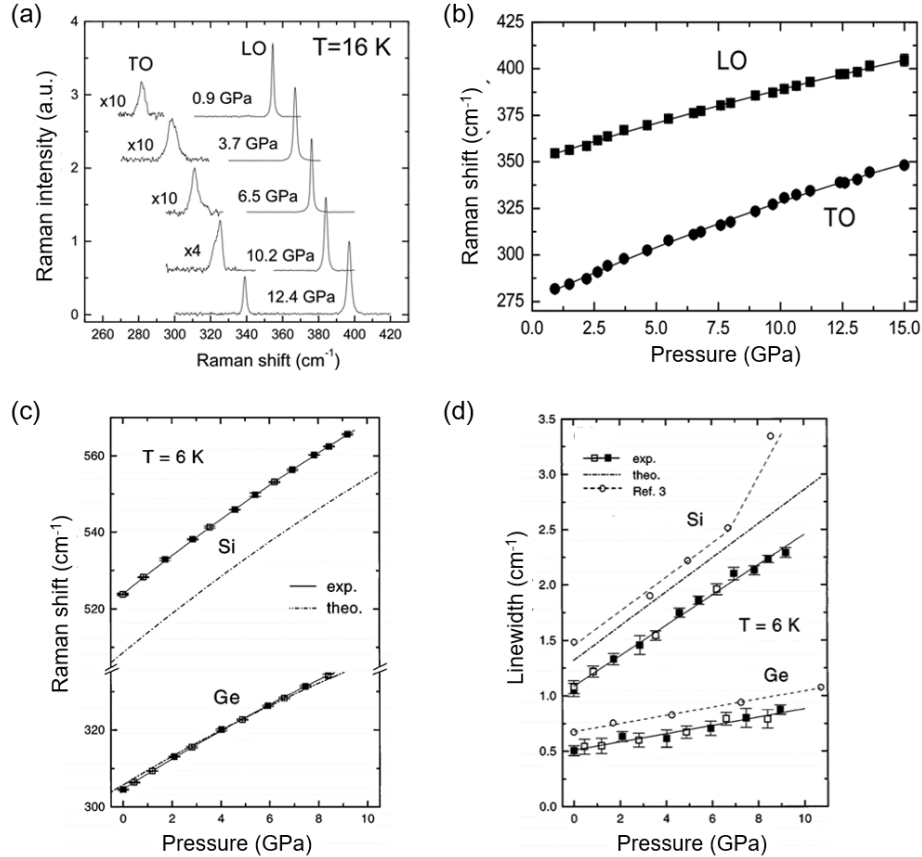

**Figure 25** Pressure-induced phonon frequency shifting and broadening by Raman spectra. (a) Raman spectra of ZnS at 16K for several pressures and (b) Pressure dependence of the phonon frequencies [200]. (c) Pressure dependence of phonon frequencies for Si and Ge. (c) Pressure dependence of the linewidth in Si and Ge. The solid lines represent the fitted functions. The dashed lines are calculated results [201].

In Figures 25(b) and 25(c), the solid curves are fits with the quadratic functions: $\omega_{LO} = 356.0 + 4.3P - 0.043P^2$ and $\omega_{TO} = 275.6 + 6.1P - 0.079P^2$ in ZnS [from the Raman spectra in Figure 25(a)] [200]; $\omega_{Si} = 523.88 + 5.1P - 0.062P^2$ and $\omega_{Ge} = 304.64 + 4.02P - 0.059P^2$ for Si and Ge, respectively [201]. In these fitted functions, it can be seen that the quadratic terms are all small enough to be ignored, i.e., the frequency shifts obey the linear relationship with pressure, which is the same as the temperature-induced quasi-harmonic effect in Eq. (3). As discussed above, the pressure changes the unit-cell volume only, i.e., the phonon-phonon interactions can be ignored at zero temperature. It can be found in Figure 25(d) that at the fixed temperature with 6 K, the pressure-dependent phonon linewidths, fitted by the least square functions, are follow the linear functions $\Gamma_{Si}$ = 1.08 + 0.137$P$ and $\Gamma_{Ge}$ = 0.51 + 0.037$P$. Although the linewidth magnitude is tiny to consider (without considering the involved finite temperature and the sample mosaic), it is still associated with the phonon decay channels changed by the pressure-induced phonon frequency shifts [201].

5.3 Practical evaluation of anharmonic interactions

As discussed above, the frequency shifts can be induced either by pressure or temperature. However, it becomes valuable to clarify that whether the frequency varies the same or not when it is induced by temperature and measured at constant pressure and when, for the same volume variation induced by pressure, it is measured at a constant temperature, i.e., fixed (varied) temperature and fixed (varied) pressure. In other words, whether (d$\ln v$ / d$V$)$_P$ equals (d$\ln v$ / d$V$)$_T$ ? Such problem can be investigated by the isothermal ($\gamma_{iT}$) and isobaric ($\gamma_{iP}$) mode Grüneisen parameters, which both quantify the change in frequency of a given mode to the change in volume of the unit cell $V$, expressed respectively as

$$\gamma_{iT} = \left(\frac{\partial \ln[v_i(P,T)]}{\partial \ln[V(P,T)]}\right)_T \quad , \quad \gamma_{iP} = \left(\frac{\partial \ln[v_i(P,T)]}{\partial \ln[V(P,T)]}\right)_P \tag{10}$$

The relationship between the implicit and explicit terms can be described as

$$\left(\frac{\partial v_i}{\partial T}\right)_P = -\left(\frac{\partial \ln V}{\partial T}\right)_P \times \left(\frac{\partial v_i(P,T)}{\partial \ln V}\right)_T + \left(\frac{\partial v_i(P,T)}{\partial T}\right)_V \tag{11}$$

The term ($\partial v/\partial T$)$_P$ can be determined by measuring the phonon spectrum as a function of temperature at constant pressure, while the volume-dependent implicit term [$\partial v_i (P, T)$ / $\partial P$]$_T$ can be obtained by measuring the phonon spectrum under pressure at a constant temperature. The explicit contribution of anharmonicity is

$$\left(\frac{\partial \ln v_i(P,T)}{\partial T}\right)_V = \alpha\left(\gamma_{iT} - \gamma_{iP}\right) = A \tag{12}$$

where $\alpha$ is the volume coefficient of thermal expansion, and A is the anharmonic parameter.

In the quasi-harmonic approximation, $\gamma_{iT}$ and $\gamma_{iP}$ are equal, and the anharmonic parameter A is zero. The ratio $\gamma_{iT} / \gamma_{iP}$ is called the implicit fraction [202]. Thus, the explicit anharmonicity, determined by the thermal population of vibrational levels, can be successfully extracted by experiments under external temperature and pressure fields. The temperature dependence of the different implicit and explicit contributions is defined as [203,204]

$$v(T) = v_0 + \Delta_1(T) + \Delta_2(T) \tag{13}$$

The implicit frequency shift can be written as [64]

$$\Delta_1(T) = -v_0(kT/16)[\Phi^{(2)}/\Phi^{(3)}] \tag{14}$$

and the explicit frequency shift can be written as [205]

$$\Delta_2(T) = \Gamma(T) = C\left[1 + N(x_1) + N(x_2)\right] + D\left[1 + 3N(x_3) + 3N(x_3)^2\right] \tag{15}$$

where C and D terms correspond to a three-phonon and a four-phonon process, respectively. $N(x)$ is the Bose-Einstein population factors of the different interacting phonons, and $x_i$ is the index of the phonon mode ($\omega_{\lambda_i}$).

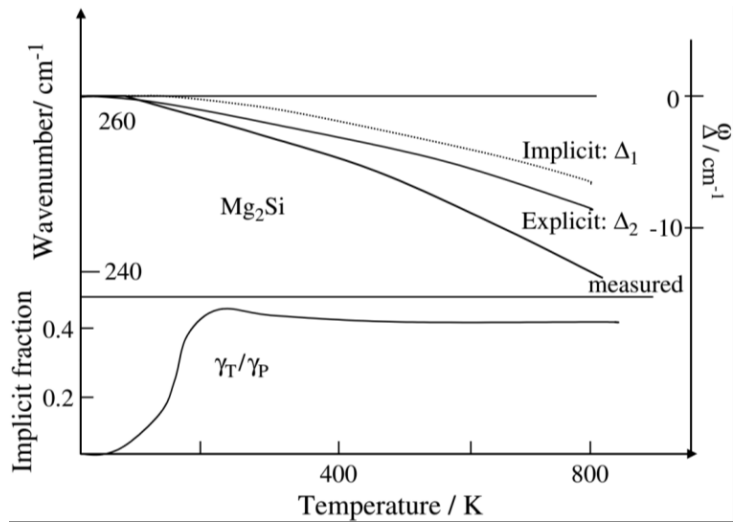

**Figure 26** Temperature-dependent Raman band of $Mg_2Si$ [202]. The implicit $\Delta_1$ and explicit $\Delta_2$ contributions are sketched along with the implicit fraction $\gamma_T/\gamma_P$ of $\gamma_T$ and $\gamma_P$.

In Figure 26, for ionic crystals, the frequency shift for T ⩽ 100 K is dominantly driven by the explicit ($\Delta_2$) anharmonicity, while for T ⩾ 300 K, the volume expansion ($\Delta_1$) is dominant [202]. Additionally, the $\gamma_T/\gamma_P$ remains constant indicates a comparable contribution from volume expansion and anharmonicity above 300 K. Such behavior is the opposite in covalent crystals [206], in which the expansion coefficient is much smaller. The temperature and pressure-induced thermodynamic properties make it possible to decouple the quasi-harmonic and anharmonic effects experimentally. Based on the description above, such decoupling can be obtained not only by the difference between the isothermal $\gamma_{iT}$ and the isobaric $\gamma_{iP}$ mode Grüneisen parameters, but also by the corresponding heat capacity, thermal expansion coefficient, and phonon entropy, etc. [4,5,74,206].

**6 Examples of phonon anharmonicity in advanced materials**

As discussed above, anharmonic phonons involve vibrations of atoms with large amplitudes. Such anomalous lattice vibration often determines the thermodynamics (e.g., thermal expansion and thermal conductivity), elastic, dielectric properties, and crystal structure of materials. In recent decades, a number of novel phenomena have been observed induced by the phonon anharmonic effects, such as ferroelectric phase transition [207,208], negative thermal expansion [31,209], ultralow thermal conductivity [32,35], making the concept of "anharmonic engineering" ever-increasingly valued by scientists. In this section, some applications of phonon anharmonicity to determine the novel physical phenomena in the related advanced materials will be introduced.

6.1 Negative thermal expansion materials

Thermal expansion is quantified by the relative change in volume, $\Delta V/V$, for a given change of the temperature $\Delta T$. The thermal expansion coefficient [210]

$$\alpha_V = \frac{1}{V}\left(\frac{\partial V}{\partial T}\right)_P \tag{16}$$

equals to zero in the harmonic approximation (*P* is the pressure and *T* is the temperature). This can be easily derived from the Gibbs distribution for the potential energy [211], which is quadratic in the atomic displacements $\boldsymbol{\mu}_j$. The thermal expansion coefficient is a positive quantity in the usual case when a positive change in temperature results in an increase in volume. However, in the case of negative thermal expansion (NTE), the volume change is negatively related to the elevated temperature. The comparison with the positive thermal expansion and NTE is shown in Figure 27 [212,213], in which one can see the volume decreased with the increased temperature in the NTE materials.

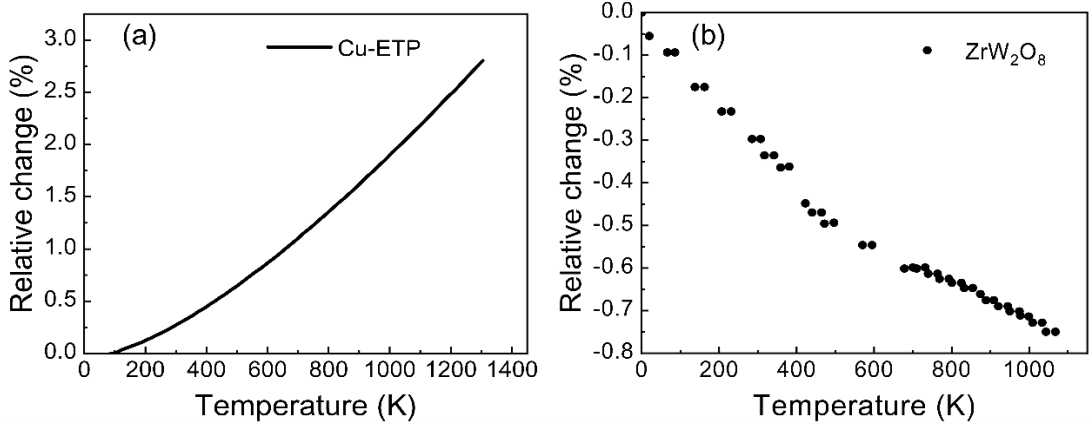

**Figure 27** Example data for the relative change in volume with temperature for the representative positive thermal expansion of Cu (a) [212] and NTE of $ZrW_2O_8$ (b) [213].

Even though the thermal expansion is an inherent process that comes from the anharmonic effect, one can employ the Grüneisen theory, phonon frequencies change depending on the volume (quasi-harmonic approximation), to explain the NTE behavior. In a crystal, we have a great number of normal phonon modes, so one defines a set of mode Grüneisen parameters:

$$\gamma_i = -\frac{V}{\omega_i}\frac{\partial \omega_i}{\partial V} = -\frac{\partial \ln \omega_i}{\partial \ln V} \qquad (17)$$

These are combined to give the mean Grüneisen parameter:

$$\overline{\gamma}(T) = \frac{1}{C_V(T)}\sum_i c_i \gamma_i \qquad (18)$$

where $C_V(T)$ is the heat capacity and $c_i$ the contribution of each phonon to the overall heat capacity, and thus, the volume thermal expansion for a crystal can be expressed as

$$\alpha_V = \frac{\bar{\gamma}(T)C_V(T)}{VB_T} \tag{19}$$

where $B_T = -V(\partial P/\partial V)_T$ is the isothermal bulk modulus. Since $C_V$, $B$ and $V$ are necessarily positive for materials in thermodynamic equilibrium, the mode Grüneisen parameters are the critical quantities for whether thermal expansion is positive or negative. It is worth noting that if we anticipate that the Grüneisen theory applies to NTE, it is required that the overall Grüneisen parameter is a negative value, not just one specific mode Grüneisen parameter.

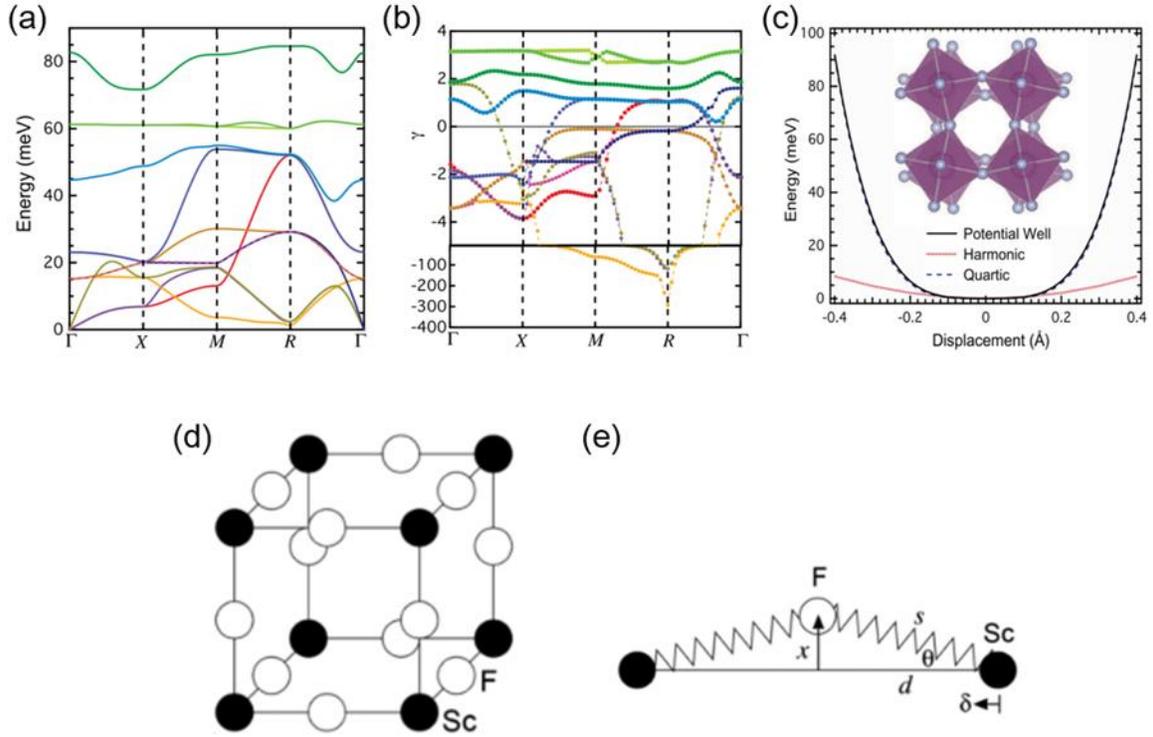

**Figure 28** Phonon properties of $ScF_3$ [112]. (a) Calculated phonon dispersions of $ScF_3$. (b) The mode Grüneisen parameters were calculated with the quasi-harmonic approximation. (c) Frozen phonon potential of the R4+ mode, and the quadratic (harmonic) and quartic fits. (d) Crystal structure of $ScF_3$. (e) Geometry and variables for the mechanical model. $x$ represents the position of F atom, $d$ half of the lattice parameter, $s$ elongations of the springs, and $\theta$ the angle between the springs and the lattice.

Here, we introduce another recent work of $ScF_3$ as an example to explain how phonon anharmonicity acts on NTE behavior. In Figure 28 [112], the low-energy rigid unit modes (RUM), $ScF_6$ octahedra pivot about corner-shared F atoms [see the crystal structure in Figure 28(d)], have ultralow phonon energies [Figure 28(a)] and large negative Grüneisen parameters [Figure 28(b)], such as at *M* and *R* points. The lowest energy mode at R point exhibits a quartic potential [Figure 28(c)], which suggests the quasi-harmonic is not reliable here, showing a strong anharmonic effect. The simple mechanical model in Figures 28(d) and 28(e) [112], describing the transverse motion of F atoms, helps to explain the inherent relationship between phonon anharmonicity and NTE. For a static state ($T = 0$), the two Sc atoms and the F atom ($x=0$) are positioned in equilibrium. In this case, there is no net force on the atoms. When the F atom starts moving (assuming the Sc atoms are fixed at their lattice positions), the transverse restoring force on the F atom mostly depends on the elongations of the springs (*s*). It goes as $1-\cos\theta$ times the resolved transverse force, giving a transverse restoring force going as $x^3$. In this case, the total potential of the F atom is $U_t = kx^4 / 4d^2$, which is consistent with the quartic potential of the R4+ mode.

NTE exists in an extensive range of materials [112,209,214], such as zirconium tungstate [213], zirconium vanadate [215], scandium tungstate [216], most of which can be described in terms of networks, or frameworks, of coordination polyhedra (e.g., perovskite or perovskite-like). Due to their novel structures, the origin of the NTE is closely related to the phonon anharmonicity. With the NTE behaviors, a number of potential applications are available for these materials [217], for example, producing composites with near-zero thermal expansion for applications where rapid temperature changes limit the performance.

6.2 Ferroelectrics and soft mode

Ferroelectric materials have garnered much attention owing to their remarkable properties and the potential applications [218], such as storages [219], transducers [220], and detectors [221]. For the displacive ferroelectrics, its phase transition is directly influenced by the soft anharmonic phonon mode [222-224]. The soft mode has a low frequency resulting from the interplay between the local restoring force and the long-range dipole interaction [225]. Due to the unique temperature-dependent damping and softening behavior, the soft mode usually behaves anharmonically in the system. In this section, the soft mode and displacive phase transitions will be briefly reviewed.

The soft mode theory of phase transitions was firstly proposed by Cochran [226] and Anderson [227]. Typically, a soft mode is a low-frequency transverse optic phonon mode in the high-temperature symmetric phase at zone center, whose frequency decreases on cooling until it falls to zero [222]. At Curie temperature, the crystal is unstable and undergoes a phase transition to a lower-symmetry phase, by freezing the atomic vibration of the soft mode. In a microscopic view, the ferroelectric phase transition originates from the relative displacement of anion and cation atoms, which gives rise to a net dipole moment of the unit cell and hence the breakdown of inversion-symmetry [Figures 29(a) and 29(b)]. Figure 29(c) shows the temperature dependence of the soft optical phonon mode of typical ferroelectric BaTiO$_3$ [222]. It can be seen that the soft mode softens with decreasing temperature. The transition on cooling also occurs as soon as any point on a phonon branch reaches zero. There is a soft mode on the low-temperature side of the transition as well, which increases in frequency on cooling, associated with the instability that occurs on heating.

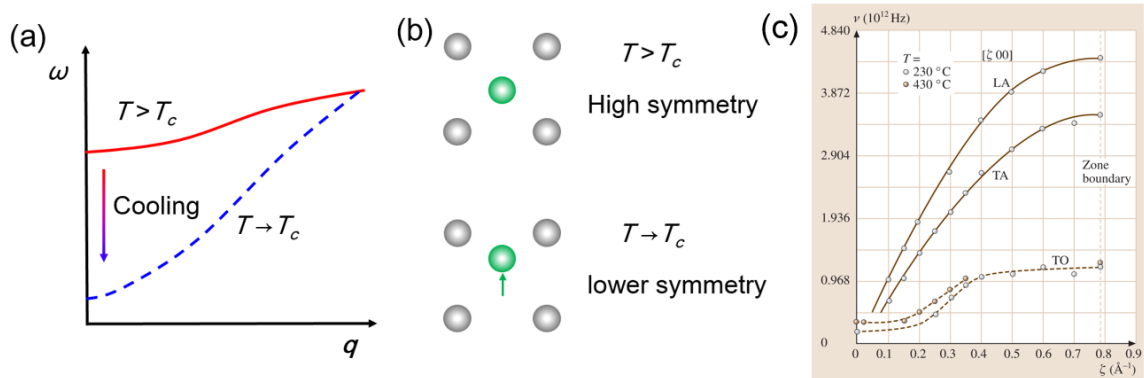

**Figure 29** Representations of a ferroelectric soft mode behavior. (a) Soft mode frequency change on cooling. (b) Corresponding atomic distortions of the soft mode [53]. (c) Temperature dependence of soft mode in BaTiO$_3$ [222].

The other way to describe the soft mode is based on the harmonic phonon dispersion calculations, in which the calculated phonon frequency in a high-temperature symmetric phase is imaginary. However, it reaches a real value in the corresponding lower-temperature and lower symmetric phase. The lower symmetry structure can be regarded as a small distortion from the higher-symmetry structure, in which the soft mode is frozen. With the soft modes from the harmonic approximation calculations, one can reach the possible ferroelectric structures by freezing the sole soft mode or the combination of soft modes, which is a standard approach in the ferroelectric community [228-230].

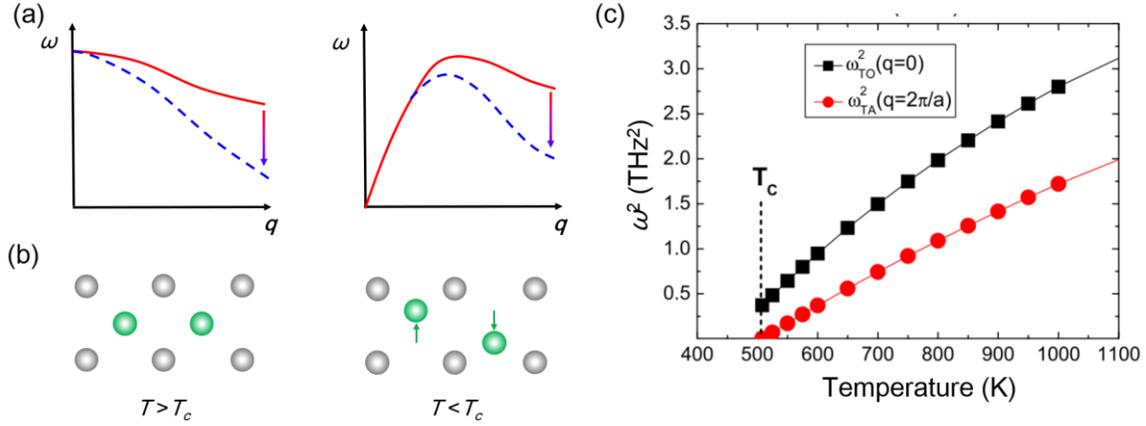

**Figure 30** Representations of zone boundary soft mode behavior. (a) Soft acoustic and optic modes on cooling. (b) Atomic displacements showing the doubling of the unit-cell and the canceling induced dipole moments. [53]. (c) Temperature dependence of the squared optic and acoustic mode frequencies of PbZrO$_3$ for $q = 0$ and $q = 2\pi / a$, respectively [224].

Another type of phase transition is related to the soft mode with wave vectors at zone boundaries, which usually emerges in anti-ferroelectric systems. In this case, the soft phonons can be either acoustic or optic modes since the distinction between the two types of mode is unclear at zone boundary [Figures 30(a) and 30(b)]. The zone boundary soft mode leads to the unit cell of the low-temperature phase doubling in one or more directions. When the soft mode at zone boundary is frozen below Curie temperature, it induces the relative displacement of anion and cation atoms in opposite directions in the neighboring unit cell, canceling the diploe moments and forming the anti-ferroelectricity. Figure 30(c) shows the temperature dependence of the squared optic (zone center) and acoustic (zone boundary) mode frequencies of PbZrO$_3$ [224]. With decreasing temperature, the acoustic softens completely at $T_C$ (507 K), driving the anti-ferroelectric phase transition of PbZrO$_3$. However, the zone center transverse optical (TO) mode remains incomplete at $T_C$, indicating the incipient behavior of ferroelectricity.

There are also many other non-ferroelectric phase transitions related to soft modes, such as quartz with the soft mode at zone center [231], ferroelastic phase transition with transverse acoustic soft mode [232], and incommensurate phase transition with the soft mode at *q* wave vector between zone center and zone boundary [233]. In all these cases, they are displacive phase transitions with symmetry breaking driven by the atomic displacement and associated with the soft modes.

6.3 Thermoelectrics and low thermal conductivity

Thermoelectric materials are widely known for the direct and reversible conversion of heat to electricity. It is a most urgent and primary goal to enhance the heat-to-electricity conversion efficiency in the past few decades [234-236]. The thermoelectric figure of merit, $ZT = S^2\sigma T / (\kappa_e + \kappa_l)$, which determines the maximum efficiency of the conversion, depends on the Seebeck coefficient ($S$), the electric conductivity ($\sigma$), and the electric ($\kappa_e$) and lattice ($\kappa_l$) thermal conductivities. [235,237] Due to the complexity of interactions between the electrical parameters, tuning the lattice thermal conductivity was proposed as a powerful stage to enhance *ZT* [238]. With this concept, reducing the lattice thermal conductivity by a so-called "phonon engineering" has been increasingly studied [37,239] in many thermoelectric materials. In this section, the phonon anharmonicity and the driving mechanism of reducing the thermal conductivity will be discussed.

If only harmonic phonons are considered, i.e., neglecting the phonon-phonon interactions and phonon scattering by any defects or quasi-particles, the phonon mean free path *l* will go infinity. This means the lattice thermal conductivity will be infinitely large for perfect, isotopically homogeneous, insulating crystals. It is the anharmonic effect that induces the finite thermal conductivity in materials. The lattice thermal conductivity is determined by the phonon group velocity $v_\lambda$ and phonon scattering rate $\tau_\lambda$:

$$\kappa_l = \frac{1}{3V}\sum_\lambda C_\lambda v_\lambda^2 \tau_\lambda \qquad (20)$$

The scattering rate in the mean free path $l = v_\lambda \tau_\lambda$ is inversely related to the phonon damping $\Gamma$ in Eq. (6), $\tau_\lambda^{-1} = 2\Gamma_\lambda$. However, from a continuum medium theory, any process of phonon-phonon interaction cannot lead to a finite thermal conductivity since the momentum is conserved at any individual interaction, and any redistribution of the momenta among the phonons will not change the energy current of the phonon gas as a whole. In crystals, Umklapp processes are possible when the momentum is conserved with the accuracy of some nonzero reciprocal lattice vector $q$, and only these processes will lead to the relaxation of the energy current and the finite $\kappa_l$ [240].

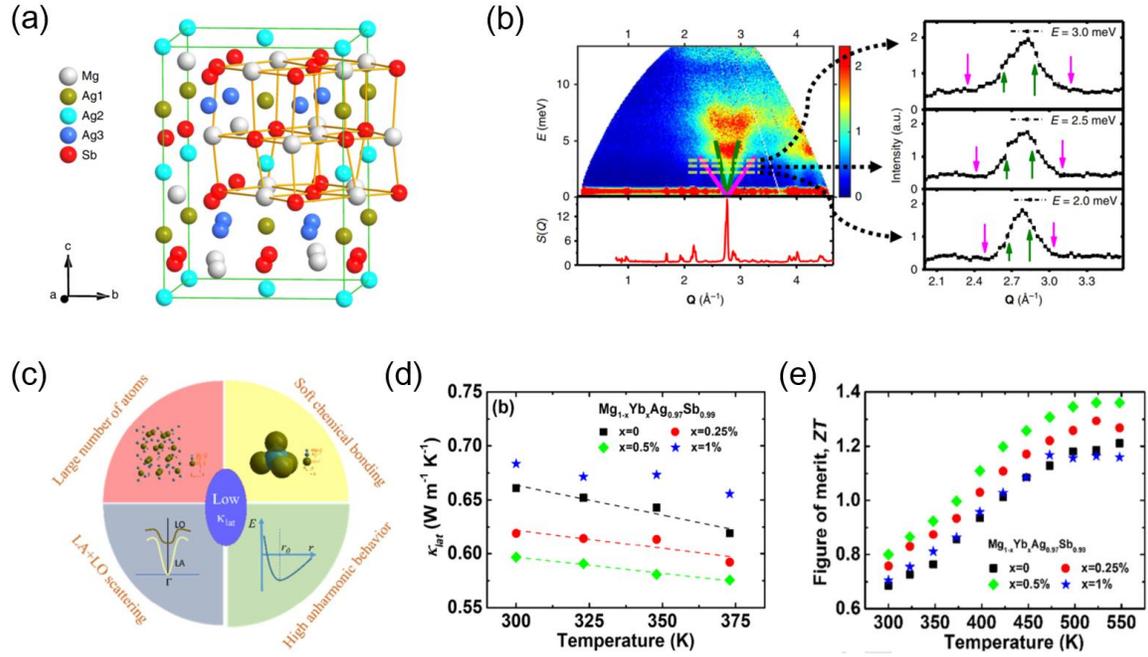

**Figure 31** Relationship between phonon anharmonicity and *ZT* in *α*-MgAgSb. (a) Crystal structure of *α*-MgAgSb with the distorted Mg–Sb rocksalt sublattice, where half of the Mg–Sb distorted cubes are filled with Ag atoms. (b) Dynamic (upper) and static (lower) structure factor measured by neutron scattering (left panel). The magenta and green lines are calculated transverse and longitudinal phonons, respectively. The energy-cut data at 2.0, 2.5, and 3.0 meV are shown in the right panel. (c) Schematic illustration of the microscopic origin of low lattice thermal conductivity for *α*-MgAgSb. (d) and (e) show the lattice thermal conductivity and *ZT* values of Yb doped *α*-MgAgSb, respectively. (a), (b), and (c) are figures from ref. [244], while (d), (e), and (f) are figures from ref. [242].

Here, we introduce a recent work of *α*-MgAgSb, a new type of promising thermoelectric material [241-243], to show the relationship between phonon anharmonicity and *ZT* (Figure 31) [242, 244]. In the primitive cell of *α*-MgAgSb, there are a large number of atoms (24 atoms), giving a very complex phonon dispersion. Mg and Sb atoms jointly form the distorted rocksalt sublattice, while Ag atoms are filled in the distorted cubes [Figure 31(a)], leading to the intrinsic weak Ag-Sb bonding. On the one hand, the TA phonons are almost fully scattered by the intrinsic distorted rocksalt sublattice [Figure 31(b)], making longitudinal acoustic (LA) phonons the primary heat carrier. However, on the other hand, the slope of the LA is abruptly suppressed by the strong anharmonic longitudinal optical phonon, known as the "avoid-crossing" interaction, which further blocks the heat transport [242]. Such unique properties make $\alpha$-MgAgSb a highly anharmonic system [Figure 31(c)], which leads to its intrinsic ultralow thermal conductivity [Figure 31(d)] via suppressing the group velocities and increasing the phonon scattering rates, and thus the high *ZT* [Figure 31(e)].

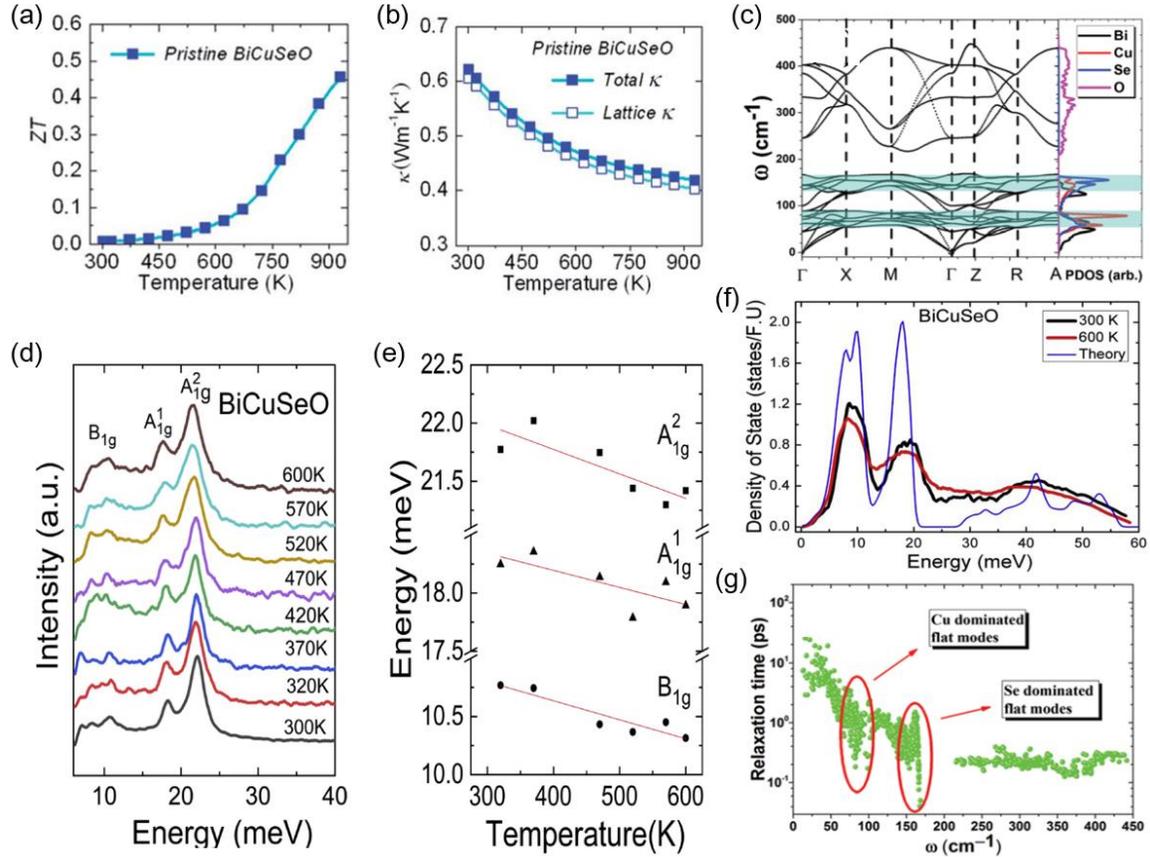

**Figure 32** Ultralow thermal conductivity induced by phonon anharmonicity in thermoelectric BiCuSeO. (a) and (b) are the measured *ZT* and thermal conductivity of BiCuSeO [245]. (c) Phonon dispersion and the DOS of BiCuSeO. The shadowed regions indicate the sharp peaks in the phonon DOS [246]. (d) and (e) are the temperature-dependence of the Raman spectra and the energy softening of three Raman-active modes [247]. (f) Comparison between the measured phonon DOS and the first-principles calculation [247]. (g) Frequency-dependence of the phonon relaxation time [246].

Recently, BiCuSeO has been attracted ever-increasing attention as a promising thermoelectric material, which mainly attribute to high *ZT* value [Figure 32(a)] dominated by the ultralow thermal conductivity [Figure 32(b)] [245]. Such low thermal conductivity can be traced to the novel phonon dispersion relations of BiCuSeO, as shown in Figure 32(c) [246]. In the low-frequency region, there exists two flat phonon bands, which correspond to the two sharp peaks in the phonon DOS [Figure 32(f)]. Based on the temperature-dependence measurements by Raman spectrometer [Figure 32(d)], it could be found that the low-frequency modes ($B_{1g}$, $A_{1g}^1$, and $A_{1g}^2$) soften with the increasing temperature [Figure 32(e)], indicating strong anharmonicity here [247]. These flat low-frequency anharmonic phonons will scatter the acoustic phonons (heat carriers) strongly, leading to relatively low relaxation times of these phonons [Figure 32(g)], and thus reduce the thermal conductivity of BiCuSeO.

In addition to the two examples mentioned in this section, other thermoelectric materials that are related to the phonon anharmonicity are discussed above as well, such as the SnSe, PbTe, $Cu_3SbSe_3$, and $Ba_8Ga_{16}Sn_{30}$, although the relationships between phonon anharmonicity and thermoelectric properties are not emphasized in these systems. For thermoelectric materials, phonon anharmonicity is a significant topic because it helps to reduce the thermal conductivity and thus enhances the *ZT* value. There are many strategies to reduce the thermal conductivity, such as "Phonon-glass Electron-crystal" [248], "Phonon-liquid Electron-crystal" [249], and lowering the dimensions [250]. Several classes of thermoelectric materials are currently under investigation, including complex chalcogenides [251], skutterudites [252], half-Heusler alloys [253], and metal oxides [254], etc. Besides thermoelectric materials that required low thermal conductivity, there are also varieties of applications in materials related to the low thermal conductivities, such as thermal barrier coating, thermal insulators, and heat sinks.

6.4 Other materials

The present review mainly focuses on phonon anharmonicity and the dominated phenomena in advanced materials. However, there are also materials closely related to phonon-particles interactions, such as ferromagnetics [255], multiferroics [256], superconductors [257], charge density wave materials [258], and some Weyl semi-metals [259]. In these cases, phonons can be scattered by electrons or spins, and the driving mechanism is known as electron-phonon or spin-phonon coupling. For example, the electron-phonon coupling drives the formation of the electron pairs responsible for superconductivity [257,260]; the spin-phonon coupling drives multiphase transitions in perovskite oxides [261]. For such strong multi-body interactions involving phonons, one can treat them as non-harmonic behaviors [4] beyond the scope of the anharmonic effects. The interested readers can find more details in the cited works.

## 7. Summary and Future Outlook

Anharmonicity is, for a long time, considered as a perturbation from the ideal harmonic ground state. With the development of anharmonic phonon theories, modern experimental techniques and computational methodologies, phonon anharmonicity, and many novel thermodynamic properties of materials are well investigated. In this review, we summarized the recent progress of the phonon anharmonicity, including the fundamental theory, characterization techniques with both experiments and computational simulations, the intrinsic and extrinsic-induced origin of anharmonicity, and the related novel phenomena induced by phonon anharmonicity in advanced materials. Although great progress has been made in the study of phonon anharmonicity in recent years, it is still a complicated, challenging, but exciting research field. The challenge arises because one should consider how phonons interact with other phonons or with other particles (or quasi-particles), while it is exciting because this interaction induces many novel properties, and the anharmonic engineering emerges to broaden its potential applications. Phonon anharmonicity attracts more and more attention, and the following aspects that are still rarely concerned may be potentially promising in the future:

(1) Anharmonic phonon interactions with other particles. In some systems, one should consider not only how phonons interact with other phonons but also with other particles, such as electron-phonon coupling [260,262], spin-phonon coupling [263,264], and phonon-spin-orbit coupling [265,266]. These complicated correlations will jointly together induce many novel properties. However, since the particles are sensitive to the crystal field and thus have a significant impact on the atomic vibrations of lattice, it is a challenge to distinguish or extract how much the novel property comes from the contribution of phonon anharmonicity, i.e., the mechanism of the origin of phonon anharmonicity in these systems are difficult to clarify. Such an issue is important because it not only helps to have in-depth understanding of the microscopic mechanism, but also helps to develop the anharmonic engineering techniques in the application of these novel properties.

(2) Quantifying the phonon anharmonicity. To our knowledge, most of the mentioned works above are focus on judging whether the phonon anharmonicity exists or not, or estimating the strength of phonon anharmonicity in the system. However, due to the experimental and simulative difficulty, there is not yet a common criterion for quantifying the anharmonicity of a real system, which is of importance for the anharmonic engineering. Therefore, it is necessary to develop experimental techniques and effective simulation approach to quantify the phonon anharmonicity.

(3) Developing the anharmonic engineering techniques. Up to now, the research on phonon anharmonicity mainly focuses on its characterization, mechanism, and its driving novel properties. However, little attention has been paid to the anharmonic engineering, i.e., utilizing anharmonicity to tune or optimize the phonon-related performance of materials. Therefore, the approaches and techniques of anharmonic engineering are in urgent need for their applications.

**Acknowledgment**

This work is supported by National Natural Science Foundation of China with Grant No. 11572040, the Beijing Natural Science Foundation (Z190011) and Beijing Institute of Technology Research Fund Program for Young Scholars. Bin Wei thanks the Doctoral Foundation of Henan Polytechnic University (Natural Science). Chen Li thanks the Initial Complement of University of California, Riverside.